\input harvmac
\input amssym.def
\input amssym
\baselineskip 14pt
\parskip 6pt

\def \d{{\rm d}}
\def \de{\delta}

\def \si{\sigma}

\def \pr{\partial}
\def \d{{\rm d}}

\def \bK{{\bar K}}

\def \bz{{\bar z}}
\def \bh{{\bar h}}
\def \bl{{\lambda}}

\def \l{\big \langle}
\def \r{\big \rangle}

\def \vep{\varepsilon}
\def \half{{\textstyle {1 \over 2}}}

\def \quar{{\textstyle {1 \over 4}}}
\def \ts{\textstyle}

\def \D{{\cal D}}
\def \E{{\cal E}}
\def \F{{\cal F}}

\def \H{{\cal H}}
\def \I{{\cal I}}

\def \O{{\cal O}}

\def\bx{{\bar x}}
\def\bz{{\bar z}}
\def\bh{{\bar h}}

\def\blam{{\bar \lambda}}
\def\tD{{\tilde D}}
\def\tDelta{{\tilde \Delta}}
\def\tF{{\tilde \F}}
\def \vphi{{\varphi}}
\def\limu#1{\mathrel{\mathop{\sim}\limits_{\scriptscriptstyle{#1}}}}
\def\toinf#1{\mathrel{\mathop{\longrightarrow}\limits_{\scriptstyle{#1}}}}

\font\ninerm=cmr9 \font\ninesy=cmsy9
\font\eightrm=cmr8 \font\sixrm=cmr6
\font\eighti=cmmi8 \font\sixi=cmmi6
\font\eightsy=cmsy8 \font\sixsy=cmsy6
\font\eightbf=cmbx8 \font\sixbf=cmbx6
\font\eightit=cmti8
\def\eightpoint{\def\rm{\fam0\eightrm}
  \textfont0=\eightrm \scriptfont0=\sixrm \scriptscriptfont0=\fiverm
  \textfont1=\eighti  \scriptfont1=\sixi  \scriptscriptfont1=\fivei
  \textfont2=\eightsy \scriptfont2=\sixsy \scriptscriptfont2=\fivesy
  \textfont3=\tenex   \scriptfont3=\tenex \scriptscriptfont3=\tenex
  \textfont\itfam=\eightit  \def\it{\fam\itfam\eightit}%
  \textfont\bffam=\eightbf  \scriptfont\bffam=\sixbf
  \scriptscriptfont\bffam=\fivebf  \def\bf{\fam\bffam\eightbf}%
  \normalbaselineskip=9pt
  \setbox\strutbox=\hbox{\vrule height7pt depth2pt width0pt}%
  \let\big=\eightbig  \normalbaselines\rm}
\catcode`@=11 %
\def\eightbig#1{{\hbox{$\textfont0=\ninerm\textfont2=\ninesy
  \left#1\vbox to6.5pt{}\right.\n@@space$}}}
\def\vfootnote#1{\insert\footins\bgroup\eightpoint
  \interlinepenalty=\interfootnotelinepenalty
  \splittopskip=\ht\strutbox %
  \splitmaxdepth=\dp\strutbox %
  \leftskip=0pt \rightskip=0pt \spaceskip=0pt \xspaceskip=0pt
  \textindent{#1}\footstrut\futurelet\next\fo@t}
\catcode`@=12 %
\def\today{\number\day\ \ifcase\month\or January\or February\or March\or April\or May\or June\or July\or
August\or September\or October\or November\or December\fi, \number\year}

\font \bigbf=cmbx10 scaled \magstep1

\lref\FGGP{S. Ferrara, A.F. Grillo, R. Gatto and G. Parisi, Covariant Expansion
of the Conformal Four-Point Function, Nucl. Phys. B49 (1972) 77\semi
S. Ferrara, A.F. Grillo, R. Gatto and G. Parisi, Analyticity Properties and
Asymptotic Expansions of Conformal Covariant Green's Functions, Nuovo Cimento 
19A (1974) 667\semi
S. Ferrara, R. Gatto and A.F. Grillo, Properties of Partial-Wave Amplitudes
in Conformal Invariant Theories, Nuovo Cimento 26A (1975) 226.}
\lref\Ftwo{S. Ferrara, A.F. Grillo and R. Gatto, Ann. Phys. 76 (1973) 161.}
\lref\Dob{V.K. Dobrev, V.B. Petkova, S.G. Petrova and I.T. Todorov,
Dynamical Derivation of Vacuum Operator Product Expansion in Conformal Field
Theory, Phys. Rev. D13 (1976) 887.}
\lref\LR{K. Lang and W. R\"uhl, The critical $O(N)$ $\sigma$-model at 
dimensions $2 <d <4$: Fusion coefficients and anomalous dimensions, 
Nucl. Phys. {B400} (1993) 597.}
\lref\Mack{G. Mack, $D$-independent representation of Conformal Field 
Theories in $D$ dimensions via transformation to auxiliary Dual Resonance 
Models, Scalar amplitudes, arXiv:0907.2407 [hep-th].}
\lref\DO{F.A. Dolan and H. Osborn, Conformal Four Point Functions and the
Operator Product Expansion, Nucl. Phys. B599 (2001) 459,  hep-th/0011040.}
\lref\DT{F.A. Dolan and H. Osborn, Conformal Partial Waves and the
Operator Product Expansion, Nucl. Phys. B678 (2004) 491,  hep-th/0309180.}
\lref\FGG{S. Ferrara, R. Gatto and A.F. Grillo, ``Conformal Algebra in 
Space-time and Operator Product Expansion", 
(Springer Tracts in Modern Physics, vol. 67) Springer (Heidelberg) 1973.}
\lref\Vilenkin{N.Ja. Vilenkin and A.U. Klimyk, ``Representation of Lie 
groups and special functions: recent advances'',
Kluwer Academic Publishers (Boston) 1995.}

\lref\Jack{H. Jack, A class of symmetric polynomials with a parameter,
Proc. Roy. Soc. Edinburgh, 69 (1970) 1.}
\lref\Pene{I. Heemskerk, J. Penedones, J. Polchinski  and J. Sully,
Holography from Conformal Field Theory, JHEP 0910:079 (2009), 
arXiv:0907.0151 [hep-th]\semi
I. Heemskerk and J. Sully,  More Holography from Conformal Field Theory,
JHEP 1009:099 (2010), arXiv:1006.0976 [hep-th].}
\lref\Rychk{ R. Rattazzi, V.S. Rychkov, Erik Tonni and A. Vichi,  
Bounding scalar operator 
dimensions in 4D CFT, JHEP 0812:031 (2008), arXiv:0807.0004 [hep-th]\semi
V.S. Rychkov and A. Vichi, Universal Constraints on Conformal Operator 
Dimensions,
Phys. Rev. D80 (2009) 045006,2009, arXiv:0905.2211 [hep-th]\semi
F. Caracciolo and V.S. Rychkov,  Rigorous Limits on the Interaction 
Strength in Quantum Field Theory,
Phys. Rev. D81 (2010) 085037, arXiv:0912.2726 [hep-th]\semi
R. Rattazzi, S. Rychkov and A. Vichi, Bounds in 4D Conformal Field Theories with 
Global Symmetry,
J. Phys. A44 (2011) 35402, arXiv:1009.5985 [hep-th]\semi
A. Vichi, Improved bounds for CFT's with global symmetries,  arXiv:1106:4037
[hep-th].
}
\lref\Pol{D. Poland and D. Simmons-Duffin, Bounds on 4D Conformal and
Superconformal Field Theories,  arXiv:1009.2087 [hep-th].}

\lref\KooR{T. Koornwinder, Two-variable Analogues of the Classical 
Orthogonal Polynomials, in Theory and Applications of Special Functions, 
ed. R. Askey, Academic Press, New York (1975).}
\lref\ortho{C.F. Dunkl and Y. Xu, Orthogonal Polynomials of Several 
Variables, Cambridge University Press, Cambridge (2001).}

\lref\NO{M. Nirschl and H. Osborn, {Superconformal Ward Identities and
their Solution}, Nucl. Phys. B711 (2005) 409, hep-th/0407060.}

\lref\ASok{F.A. Dolan, L. Gallot and E. Sokatchev, {On Four-point Functions
of $\half$-BPS operators in General Dimensions}, JHEP 0409 (2004) 056,
hep-th/0405180.}

\lref\Ko{T. H. Koornwinder, Orthogonal polynomials in two variables which 
are eigenfunctions of two algebraically independent partial differential 
operators. I, II, Nederl. Akad. Wetensch. Proc.
Ser. A 77 = Indag. Math. 36 (1974), 48, 59.}
\lref\Koo{T. Koornwinder and I. Sprinkhuizen-Kuyper, Generalized Power
Series Expansions for a Class of Orthogonal Polynomials in Two Variables,
Siam J. on Mathematical Analysis, 9 (1978) 457.}
\lref\Vretare{L. Vretare, Formulas for Elementary Spherical Functions and
Generalized Jacobi Polynomials, Siam J. on Mathematical Analysis, 15 (1984)
805.}
\lref\Koop{T. H. Koornwinder, Lowering and raising operators for some 
special orthogonal polynomials,  arXiv:math/0505378.}

\lref\Bate{H. Bateman, The solution of partial differential equations by 
means of definite integrals, Proceedings of the London Mathematical 
Society 1 (1904) 451.}

\lref\Spec{G.E. Andrews, R.Askey and R. Roy, ``Special Functions", Cambridge
University Press (Cambridge) 1999.}

\lref\Pet{H. Osborn and A. Petkou, Implications of conformal invariance in 
field theories for general dimensions,   Ann. Phys. {231} (1994) 311; 
hep-th/9307010.}

\lref\Syman{K. Symanzik, On Calculations in conformal invariant field 
theories, Lett. al Nuovo Cimento 3 (1972) 734.}

\lref\pene{J. Penedones, Writing CFT correlation functions as AdS 
scattering amplitudes, JHEP 1103:025 (2011), arXiv:1011.1485 [hep-th].}

\lref\penet{A.L. Fitzpatrick, J. Kaplan, J. Penedones, S. Raju and
B.C. van Rees, A Natural Language for AdS/CFT Correlators, JHEP 1111 (2011) 095,
arXiv:1107.1499 [hep-th].}

\lref\paulos{M.F. Paulos, 
Towards Feynman rules for Mellin amplitudes in AdS/CFT, JHEP 1110 (2011) 074,
arXiv:1107.1504 [hep-th].}

{\nopagenumbers
\rightline{DAMTP/11-64}
\rightline{arXiv:1108.6194[hep-th]}
\rightline{\today}
\vskip 1.5truecm
\centerline {\bigbf Conformal Partial Waves:}
\vskip  6pt
\centerline {\bigbf Further Mathematical Results}
\vskip 2.0 true cm
\centerline {F.A. Dolan${}^*$ and H. Osborn${}^\dagger$}

\vskip 12pt
\centerline {${}^*$Department of Physics, University of Crete,} 
\centerline {GR-71003, Heraklion, Greece}
\vskip 6pt

\centerline {${}^\dagger$Department of Applied Mathematics and Theoretical Physics,}
\centerline {Wilberforce Road, Cambridge, CB3 0WA, England}
\vskip 1.5 true cm

{\eightpoint
\parindent 1.5cm{

{\narrower\smallskip\parindent 0pt
Further results for conformal partial waves for  four point functions for 
conformal primary scalar fields in conformally invariant theories are 
obtained. They are defined as eigenfunctions of the differential Casimir 
operators for the conformal group acting on two variable functions subject 
to appropriate  boundary conditions. As well as the scale dimension $\Delta$ 
and spin $\ell$ the conformal partial waves depend on two parameters $a,b$ 
related to the dimensions of the operators in the four point function. 
Expressions for the Mellin transform of conformal partial waves are obtained
in terms of polynomials of the Mellin transform variables given in terms
of finite sums. Differential  operators which change $a,b$ by $\pm 1$, 
shift the dimension $d$ by $\pm 2$ and also change $\Delta,\ell$ are found. 
Previous results for $d=2,4,6$ are recovered. The trivial case of $d=1$ 
and also $d=3$ are also discussed. For $d=3$ formulae for the conformal 
partial waves in some restricted cases as a single variable integral 
representation based on the Bateman transform are found.

PACS no: 11.25.Hf

Keywords: Conformal field theory, Operator product expansion, Four point
function

\narrower}}}
\medskip

\parindent 1.5cm{

{\narrower\smallskip\parindent 0pt
{\it Very shortly after the completion of this paper Francis Dolan died,
may this be a small contribution to his memory.

\narrower}}}

\vfill
\line{\hskip0.2cm email:${}^\dagger${\tt ho@damtp.cam.ac.uk}}
\hfill}


\eject

\pageno=1
\newsec{Introduction}

In conformal field theories the correlation functions of conformal primary 
operators $\phi_i$ satisfy non trivial identities obtained by expanding various 
pairs of operators $\phi_i\phi_j$ appearing in the correlation function using 
the operator product expansion. Assuming that in such correlation functions 
$\phi_i(x_i)\phi_j(x_j)$ has a convergent operator product expansion in terms of 
a complete set of of operators with differing spins and scale dimensions then 
it is natural to formulate bootstrap equations relating the expansions for each 
pair $i,j$. Individual contributions to each expansion are identified as 
conformal partial waves if they include the contributions in the operator 
product expansion of a conformal primary operator $\O_\Delta^{(\ell)}$, with 
scale dimension $\Delta$ and spin $\ell$, and also all its descendants formed 
by the action of derivatives on
$\O_\Delta^{(\ell)}$. To find ways of making the bootstrap equations tractable 
it is important to have succinct mathematical expressions for the conformal 
partial waves.

For four point correlation functions the conformal partial waves $G_\Delta^{(\ell)}$
are functions of two conformal invariants $u,v$. 
They can be determined by requiring them to be eigenfunctions of the Casimir 
operators of the conformal group, subject to appropriate boundary conditions, 
with eigenvalues given in terms of $\Delta,\ell$.  The conformal group is 
non compact so that $\Delta$ is allowed to vary continuously, although subject 
to restrictions to ensure unitarity. From 
the 1970's onwards, when the conformal bootstrap was first proposed, various 
formulae for the conformal partial waves $G_\Delta^{(\ell)}$ have been derived 
although these are  often rather involved \refs{\FGGP,\Dob,\LR}. More recently 
expressions for conformal partial waves as functions of $u,v$ have been 
obtained in terms of Mellin-Barnes integral  representations 
\refs{\Mack,\pene,\penet,\paulos}.

Acting on $G_\Delta^{(\ell)}(u,v)$ the quadratic Casimir
reduces to a linear second order operator in $u,v$. For $\ell=0$ solutions
may be obtained as a double power series in $u,1-v$ although there is no
simple generalisation for $\ell>0$. With a Mellin-Barnes representation
where the conformal partial wave is expressed as a $s,t$ integral over $u^s v^t$
the eigenvalue equation reduces to a six term difference equation involving
the two independent integration variables $s,t$ \penet. As shown here
in the particular case of  leading twist, essentially for dimension $d$
when $\Delta = \ell+d-2$,
the conformal partial wave is annihilated by an additional second order
differential operator. This renders finding explicit solutions in this
case, by various techniques, much more tractable. The leading twist case
is particularly relevant in superconformal theories when it corresponds to
the contribution of semi-short operators.

In two and four dimensions, with  extensions possible for any even dimension, 
quite simple results have been found which have been found useful in bootstrap 
applications \refs{\Rychk,\Pene,\Pol}. These are described in two previous 
papers \refs{\DO,\DT}. 
The essential step to achieve a relatively simple form for the conformal 
partial waves is to express them in terms of symmetric functions of two 
variables $x,\bx$, which are determined by the conformal invariants $u,v$. 
In terms of $x,\bx$ the eigenvalue equation simplifies and the 
conformal partial waves involve just products of ordinary 
hypergeometric functions, depending on three parameters, of $x$ and $\bx$.
The degree of complication is independent of $\ell$.
In the leading twist case the conformal partial wave reduces to a single
variable function. In terms of the variables $x,\bx$ the  additional second
order operator for leading twist is a reduction of the $d$-dimensional
wave operator, and the simplifications for even dimensions are a reflection
of Huygens principle.
 
In many respects results for conformal partial waves are a non compact 
analogue to expressions for two variable harmonic polynomials, for which 
there is an extensive literature \refs{\Ko,\Koo,\KooR,\Koop} and which also 
extends to consideration of symmetric functions of $p$ variables for any $p$ 
\refs{\Vretare,\ortho}.
For $p=1$ the harmonic polynomials of interest are just the classical 
Jacobi polynomial $P_n^{(\alpha,\beta)}(y)$ which may be defined by requiring
them to be orthogonal for $n=0,1,2,\dots$ with respect to the 
measure $\d y\, (1-y)^\alpha(1+y)^\beta$ for $y \in [-1,1]$. 
Their symmetric two variable extensions defined in \Ko\ 
$P_{nm}^{(\alpha,\beta,\gamma)}(y+{\bar y},y{\bar y})$
are orthogonal with respect to the measure
$\d y\, \d {\bar y} \; \big ( (1-y)(1- {\bar y})\big ){}^\alpha
\big ( (1+y)(1+ {\bar y})\big ){}^\beta \, |y- {\bar y} |^{2\gamma+1}$.
For $\gamma= \pm \half$ there is an explicit construction in terms of
the single variable  Jacobi polynomials which is
identical to the formulae for dimension $d=2,4$ conformal partial 
waves in terms of hypergeometric functions. In general the correspondence
requires $2 \gamma \to d - 3$. For any $\gamma$ in \refs{\Ko,\Koop} it was 
shown that the two  variable harmonic polynomials satisfy various recurrence 
relations for which there may be expected to be equivalent results in the 
non compact case. Previous results \DT\ gave partial support for 
this and we endeavour to derive in this paper expressions for derivative
operators relating conformal partial waves.

In section 2 of this paper we define conformal partial waves in terms of
eigenfunctions of the quadratic Casimir of the conformal group.
Single variable functions $g_p(a,b;x)$ which play an essential role in the
construction of conformal partial waves are introduced. These are
hypergeometric functions, up to an additional power $x^p$, and  are the 
non compact analogs of the Jacobi polynomials $P_n^{(\alpha,\beta)}(y)$. 
The equations for the conformal partial waves are simplified
by introducing the new variables, $x,\bx$, so that they become symmetric 
functions $F_{\lambda_1\lambda_2}(a,b;x,\bx)$ with $\lambda_{1,2}
= \half (\Delta \pm \ell)$ and $a,b$ determined by the dimensions of
of the operators appearing in the four point function.
In section 3 we describe a conformally invariant integral
which may be expressed in terms of the sum of a conformal partial wave
corresponding to the contributions arising from an operator 
$\O_\Delta^{(\ell)}$ and also its shadow 
${\widetilde {\O}}^{(\ell)}_{d-\Delta}$. This leads to a Mellin-Barnes
integral expression which reduces the problem of determining conformal
partial waves to solving a multi-term difference equation with polynomial 
solutions in the two complex variables $s,t$ of degree $\ell$.
Explicit results are given for $\ell=1,2$. More general solutions 
have an algebraic complexity which increases rapidly with $\ell$, although
results may be obtained for the leading twist case by solving a three term
difference equation.
 
In section 4 various recursion relations for conformal partial
waves in general dimensions are obtained. These involve the construction
two sets of differential operators each of which are related by considering 
commutators with the differential operator representing the quadratic Casimir. 
By including the quartic Casimir, which is a fourth order differential operator,
 these form two closed algebras which ensure that their action in each case 
 generates a finite number of conformal partial waves
with  $\Delta,\ell$ differing by $\pm 1$ and in the first example also  $a,b$
differing  by $\pm \half$.
We also construct second order differential operators which acting on
$F_{\lambda_1\lambda_2}(a,b;x,\bx)$ change $a,b$ by $\pm 1$ and
the dimension $d$ by $\pm 2$. Applying these shift operators twice
leads to eigenvalue equations which are equivalent to those given by
the quartic Casimir. In section 5
previous results found in \DT\ for $d=4,6$ are obtained from the $d=2$
result using the dimension raising operator. The conformal partial
waves are formed from two $g_p$ functions depending on $x,\bx$.
In section 6 conformal partial waves for $d=1$, when they reduce to 
just one single variable function $g_p(x)$ since kinematics require 
$x=\bx$ in this case, are found. For $d=3$ solutions for restricted
$\lambda_2$ are found using an integral representation related to the
Bateman transform for solutions of the three dimensional wave equation.

Some supplementary mathematical results are described in five appendices.
Appendix A evaluates some relevant two dimensional conformal integrals.
In appendix B we show how analogous results are obtained for the two
variable extensions of Jacobi polynomials and which may be mostly found
in the mathematical literature. In appendix C we show how the apparent
singularity in the six dimensional case cancels while in appendix D
we obtain some results for conformal partial wave expansions when
only leading twist operators are present. An alternative construction
for conformal partial waves based on an expansion in terms of Jack
polynomials, used previously in \DT, in described briefly in appendix E
although the general form is very complicated.

\newsec{Definitions and General Results for Conformal Partial Waves}

The four point function for four scalar primary operators $\phi_i$
with scale dimensions $\Delta_i$ has the general form
\eqn\fourp{\eqalign{
 \l & \phi_1(x_1) \, \phi_2(x_2) \,\phi_3(x_3) \,\phi_4(x_4) \r \cr &{} 
 = {1\over (x_{12}{\!}^2)^{{1\over 2}(\Delta_1 + \Delta_2)} \, 
(x_{34}{\!}^2)^{{1\over 2}(\Delta_3 + \Delta_4)}}\, \bigg ( {x_{24}{\!}^2 
\over x_{14}{\!}^2} \bigg)^{\!{1\over 2}\Delta_{12}} \bigg ({ 
{x_{14}{\!}^2} \over x_{13}{\!}^2} \bigg)^{\! {1\over 2}\Delta_{34}} 
F(u,v) \, , \cr} } 
where 
\eqn\Ddiff{ x_{ij} = x_i - x_j \, , \qquad 
\Delta_{ij} = \Delta_i - \Delta_j \, , } and $u,v$ are the two independent 
conformal invariants 
\eqn\defuv{ 
u= {x_{12}{\!}^2 \, x_{34}{\!}^2\over x_{13}{\!}^2 
\, x_{24}{\!}^2 } \, , \qquad v= { x_{14}{\!}^2 \, x_{23}{\!}^2 \over 
x_{13}{\!}^2 \, x_{24}{\!}^2 } \, . } 
The conformal partial waves 
$G_\Delta^{(\ell)}(u,v)$ are so defined so that the conformal partial wave 
expansion becomes
\eqn\part{
F(u,v) = \sum_{\Delta,\ell} a_{\Delta,\ell} \; 
G_\Delta^{(\ell)}(u,v) \, .
} 
where $G_\Delta^{(\ell)}(u,v)$ is  the  contribution of an operator  
$\O_\Delta^{(\ell)}$, with scale dimension $\Delta$ and spin $\ell$,
and its descendants to  the operator product expansion of 
$\phi_1\phi_2$ or $\phi_3\phi_4$.

The operator $\O_\Delta^{(\ell)}$ and also all its descendants are
eigenvectors of the Casimir operators formed from the generators 
$M_{AB}= - M_{BA}$, $A,B=0,1,\dots , d+1$,
of the conformal group $SO(d,2)$ acting on  $\O_\Delta^{(\ell)}$ and
which obey the algebra
\eqn\Lalg{
[M_{AB},M_{CD}] =  \eta_{AC}\, M_{BD} - \eta_{BC}\, M_{AD} - 
\eta_{AD}\, M_{BC} + \eta_{BD}\, M_{AC} \, ,
}
with $\eta_{AB} = {\rm diag.}(-1,1,\dots , 1, -1)$.
For the quadratic and quartic Casimirs constructed from $M_{AB}$ we then have
\eqn\CasO{\eqalign{
\half \, M^{AB}M_{BA} \, \O_\Delta^{(\ell)} = {}& C_{\Delta,\ell} \,
\O_\Delta^{(\ell)} \, , \cr
\half \, M^{AB}M_{BC}M^{CD}M_{DA} \, \O_\Delta^{(\ell)} = {}& D_{\Delta,\ell} \,
\O_\Delta^{(\ell)} \, , 
}
}
where
\eqn\Ceig{\eqalign{
C_{\Delta,\ell} = {}& \Delta(\Delta -d) + \ell (\ell+d-2) \, , \cr
D_{\Delta,\ell} = {}& \Delta^2(\Delta -d)^2 + \half d(d-1)\, \Delta(\Delta -d)\cr
\noalign{\vskip -2pt}
&{} + \ell^2 (\ell+d-2)^2 + \half (d-1)(d-4) \, \ell (\ell+d-2) \, .
}
}

If $M_{i\,AB}$ are the generators  acting on $\phi_i$
then conformal invariance of the correlation function requires $\sum_i M_{i\,AB}
\l \phi_1(x_1) \, \phi_2(x_2) \,\phi_3(x_3) \,\phi_4(x_4) \r = 0$. The essential
equation determining the conformal partial waves is then 
\eqn\Cas{
\big ( \half (M_1 + M_2)^{AB} (M_1+M_2)_{BA}  - C_{\Delta,\ell} \big ) 
{1\over ( x_{12}{\!}^2)^{{1\over 2}(\Delta_1 + \Delta_2)} }\, 
\bigg ( {x_{24}{\!}^2 \over x_{14}{\!}^2} \bigg)^{\!{1\over 2}\Delta_{12}} \!  
\bigg ({ {x_{14}{\!}^2} \over x_{13}{\!}^2} \bigg)^{\! {1\over 2}\Delta_{34}} 
\!\! G_\Delta^{(\ell)} (u,v) =0  \, .
}
This leads directly to the two
variable second order equation
\eqn\DGeig{
\D \, G_\Delta^{(\ell)}(u,v) = \half C_{\Delta,\ell} \, 
G_\Delta^{(\ell)}(u,v) \, ,
}
for
\eqn\LG{\eqalign{
\D = {}& (1- u- v ) {\pr \over \pr v}  \bigg (
v {\pr \over \pr v} + a+b \bigg )  +
u {\pr \over \pr u}  \bigg ( 2\, u {\pr \over \pr u} - d \bigg )   \cr
& {} - (1+u-v)  \bigg ( u {\pr \over \pr u} +  v {\pr \over \pr v}  + a \bigg )
\bigg ( u {\pr \over \pr u} +  v {\pr \over \pr v} +b  \bigg )  \, ,  \cr}
}
where
\eqn\ab{
a = - \half  \Delta_{12} \, , \qquad b = \half \Delta_{34} \, .
}

The contribution of an operator $\O_\Delta^{(\ell)} $ to the four point 
function from the operator product expansion 
 then has the form
\eqn\fourpG{\eqalign{
 \l & \phi_1(x_1) \, \phi_2(x_2) \,\phi_3(x_3) \,\phi_4(x_4) \r 
 \big |_{\phi_1\phi_2, \phi_3\phi_4 \sim \O_\Delta^{(\ell)}} \cr
&{} = {1\over (x_{12}{\!}^2)^{{1\over 2}(\Delta_1 + \Delta_2)} \, 
(x_{34}{\!}^2)^{{1\over 2}(\Delta_3 + \Delta_4)}}\, \bigg (
{x_{14}{\!}^2 \over x_{24}{\!}^2} \bigg)^{\! a} \bigg (
{x_{14}{\!}^2 \over x_{13}{\!}^2} \bigg)^{\! b} a_{\Delta,\ell} \, 
G_\Delta^{(\ell)}(u,v)  \, .  } 
}
It follows directly that the conformal partial waves $G_\Delta^{(\ell)}$ 
satisfy symmetry relations under $1\leftrightarrow 2$ and $3\leftrightarrow 4$,
\eqn\symG{
G_\Delta^{(\ell)} \big ( u/v , 1/v) 
= (-1)^\ell \, v^b G_\Delta^{(\ell)}(u,v) \, \big |_{a\to -a} 
= (-1)^\ell \, v^a G_\Delta^{(\ell)}(u,v) \big |_{b \to - b} \, ,
}
and, by considering $1 \leftrightarrow 4, \, 2 \leftrightarrow 3$, 
$G_\Delta^{(\ell)}$  is also a symmetric function of $a,b$. 

To leading order as $u\to 0 , \, v\to 1$, 
$\D(u^p (1-v)^q ) \sim \big ( p (2p-d) + q (2p+q-1)\big ) u^p (1-v)^q$
so that the relevant solutions of  \DGeig\ for conformal partial waves 
may be  determined by imposing the boundary conditions
\eqn\Gsim{ 
G_\Delta^{(\ell)}(u,v) = c_\ell \,   
u^{{1\over 2}(\Delta-\ell)}(1-v)^\ell \big ( 1 + {\rm O}(u,1-v) \big ) \, .
}
The constant $c_\ell$ in 
\Gsim\ is inessential and may be set to $1$ but it is convenient here to 
allow for more general possibilities, although we require $c_0=1$.

As previously mentioned for general dimensions $d$ and $\ell>0$ results for
$G_\Delta^{(\ell)}(u,v)$  do not have any simple form and so are not readily 
useful.  For $d=2,4$ more explicit 
expressions have been found \refs{\DO,\DT}. These involve using the different 
variables $ x,\bx$ defined in terms of $u,v$ by 
\eqn\defxz{ u = x \bx \, , \qquad v = (1-x)(1-\bx) \, . } 
In the Euclidean domain $x,\bx$ are conjugates.
With \defxz\ $G_\Delta^{(\ell)}$ becomes a symmetric function of $x,\bx$ 
\eqn\GF{ G_\Delta^{(\ell)}(u,v) = F_{\lambda_1\lambda_2}(x,\bx) = 
F_{\lambda_1\lambda_2}(\bx,x) \, , } 
for 
\eqn\lll{
\bl_1 = \half ( \Delta + \ell) \, , \quad \bl_2 = \half ( \Delta - \ell) \, , 
\qquad \ell = \bl_1 - \bl_2 = 0,1,2, \dots \, . } 
Corresponding to \Gsim\ we have the boundary conditions
\eqn\Fsim{ F_{\bl_1 \bl_2}(x,\bx) \limu{\bx\to 0,x\to 0} \cases{
c_\ell \, (x\bx)^{\lambda_2} \, x^\ell \, , \ \ &$x\bx>0 \, ,$ \cr
c_{\bl_1 -\bl_2} \, x^{\bl_1}\bx^{\bl_2} \, , \ \ & $x,\bx >0 \, ,$} } 
where the limit $\bx\to 0$ is taken first. 
The differential operator $\D$ in \LG\ becomes in terms of $x,\bx$
\eqn\defD{\eqalign{ \Delta^{(\vep)} 
\equiv \Delta^{(\vep)}(a,b) = {}& D_x(a,b) + D_\bx(a,b) \cr 
&{} + 2\vep \, {x\bx \over x-\bx} \Big ( (1-x) {\pr \over \pr x} - (1- \bx) 
{\pr \over \pr \bx} \Big ) \, , }}
for $D$ the single variable differential operator defined by
\eqn\hyper{ 
D_x(a,b) 
= x^2(1-x) {\d^2 \over \d x^2} -(a+b+1)\, x^2 {\d \over \d x} - ab\, x 
\,, }
and for
\eqn\abc{ 
\vep = \half(d-2) \, . }
Unless necessary the dependence of $\Delta^{(\vep)}$, and also $D_x$
in \hyper, on $a,b$ is suppressed when there is no ambiguity.

The eigenvalue equation \LG\ is then equivalent to
\eqn\eigF{ 
\Delta^{(\vep)} F_{\bl_1 \bl_2}(x,\bx) = c_{\lambda_1\lambda_2} \,
F_{\bl_1 \bl_2}(x,\bx) \, ,  \quad
c_{\lambda_1\lambda_2}= \bl_1 ( \bl_1  - 1 ) + \bl_2 ( \bl_2  - 1 - 2\vep )\, , 
} 
for symmetric functions $F_{\bl_1 \bl_2}(x,\bx)$ 
subject to the boundary conditions \Fsim. 
When it is desirable  to exhibit the dependence on $a,b$ or $\vep$  the 
conformal partial waves are written as  $ F_{\bl_1 \bl_2}(x,\bx) \equiv 
F^{(\vep)}{\!}_{\bl_1 \bl_2}(a,b;x,\bx) $.
The symmetry relations \symG\ by virtue of \GF\ now become
\eqna\symF
$$\eqalignno{
F_{\lambda_1\lambda_2}(a,b;x',\bx') = {}& (-1)^\ell v^b
F_{\lambda_1\lambda_2}(-a,b;x,\bx) \, , & \symF{a} \cr
= {}& (-1)^\ell v^a F_{\lambda_1\lambda_2}(a,-b;x,\bx) \, , &\symF{b} }
$$
with $v$ as in \defxz\ and
\eqn\xxp{
x' = { x \over x-1} \, , \qquad \bx' = { \bx \over \bx-1}  \, .
}

The conformal partial waves $F_{\bl_1 \bl_2}(x,\bx)$ may be regarded
as functions of the elementary symmetric functions of two variables,
$x + \bx, \, x\bx$. The limiting result given by \Fsim\ may be extended for
$x = {\rm O}(\bx)$ to the form
\eqn\Fxx{
F_{\bl_1 \bl_2}(x,\bx) \limu{x,\bx\to 0} (x\bx)^{{1 \over 2}\Delta} \, 
\big ( f(\sigma) + {\rm O}(x+\bx, \sqrt{x\bx}) \big )\, ,
}
for 
\eqn\defsi{
\sigma = {x+\bx \over 2 (x\bx)^{{1\over 2}}} \, .
}
Substituting \Fxx\ into \eigF, with the boundary conditions \Fsim, requires
\eqn\eqf{
(\sigma^2 - 1) f''(\sigma) + (1+2\vep) \, \sigma f'(\sigma) = \ell (\ell+2\vep)
f(\sigma) \, , \qquad f(\sigma) \limu{\sigma \to \infty} c_\ell (2\sigma)^\ell
\, .
}
The relevant solutions of \eqf\ are Gegenbauer polynomials 
we may then write
\eqn\fssol{
f(\sigma) = {\hat C}^{\,\vep}_\ell (\sigma)  \, ,
}
choosing an alternative normalisation so that
\eqn\CCl{
{\hat C}^{\, \vep}_\ell (\sigma) = 
{\ell! \over (2\vep)_\ell} \, C^{\,\vep}_\ell(\sigma) \, , \qquad
{\hat C}^{\, \vep}_\ell(1)=1\, .
}
In this case in \Fsim\
\eqn\cell{
c_\ell \equiv c^{(\vep)}{\!}_\ell =  {(\vep)_\ell\over (2\vep)_\ell}  \, .
}

An additional reduction to a single variable function, for any $\vep$,
may be found by considering a limit in which $\bx \to 0$ since,
using $ \Delta^{(\vep)}  \sim  D_x + \bx^2 {\pr^2 \over \pr \bx^2}
- 2\vep \, \bx {\pr \over \pr \bx}$, 
\eqn\Dqp{
\Delta^{(\vep)} \big ( \bx^q \, g(x) \big ) = \bx^q \big (
D_x g(x) + q(q -1-2\vep) g(x) \big )  + {\rm O} (\bx^{q+1} ) \, .
}
Hence the limiting behaviour \Fsim\ alternatively extends to 
\eqn\Fxsim{ 
F_{\bl_1 \bl_2}(x,\bx) \limu{\bx\to 0} 
c_{\bl_1 -\bl_2} \, \bx^{\bl_2} \, g_{\lambda_1}(x) \, , } 
so long as $g_{\lambda_1}$ is a solution of 
\eqn\defg{ 
D_x(a,b) \, g_p(a,b;x) = p(p-1) \, g_p(a,b;x) \, , }
subject to
\eqn\glim{
g_p(a,b;x) \limu{x\to 0} x^p  \, .
}
Since 
\eqn\Dsol{ 
x^{-p} D_x(a,b) \, x^p = x^2 (1-x) 
{\d^2 \over \d x^2} + x \big (2p -(2p+a+b+1) x \big ) {\d \over \d x} 
- (p+a)(p+b) \, x  + p(p-1) \, , } 
the functions $g_p$ defined by \defg\ and \glim\ may be identified
as just standard hypergeometric functions,\foot{If $g_p(a;b;x)$ is
a solution of \defg\ then so is $g_{-p+1}(a,b;x)$ but the required solution is
determined by the boundary condition \glim\ and we further assume $p\ge 0$.}
\eqn\hyper{ 
g_p(a,b;x) = x^p F(p+a,p+b;2p;x) \, .  } 
The single variable functions $g_p$ defined by 
\hyper\ are essential for constructing general expressions for 
$F_{\bl_1 \bl_2}(x,\bx)$ for any $\vep$, they play an analogous
role for non compact $SO(1,2)$ to the harmonic Jacobi polynomials 
$P_n^{(\alpha,\beta)}$ in relation to compact $SO(3)$.

There are various identities for the $g_p$ functions inherited
from those for hypergeometric functions such as
\eqn\gab{\eqalign{
g_p(a,b;x) = {}& e^{\pm i \pi p}(1-x)^{-b} g_p(-a,b;x') = 
e^{\pm i\pi p}(1-x)^{-a} g_p(a,-b;x') \cr
= {}& (1-x)^{-a-b} g_p(-a,-b;x) \, ,
}
}
where $x'$ is defined in \xxp. For $p$ non integer
the choice of factors $e^{\pm i\pi p}$ depends on which side 
of the cut with branch point at $x=0$ arising from $x^p$ the
continuation from $x\in (0,1)$ to $x'<0$ is defined.
The last equality is a reflection of the identity
\eqn\idD{
D_x(a,b) = (1-x)^{-a-b} D_x(-a,-b)\, (1-x)^{a+b}\, .
}

The result \Fxsim\ is equivalent to
\eqn\Gusim{
G_\Delta^{(\ell)}(u,v) \sim c_\ell \, 
u^{{1\over 2}(\Delta-\ell)}(1-v)^\ell 
F \big ( \half(\Delta + \ell) +a , \half(\Delta + \ell) +b ; \Delta+\ell;
1-v \big ) \quad \hbox{as} \quad u \to 0  \, ,
}
extending \Gsim.

The operator $\Delta^{(\vep)}$ also satisfies identities which are of 
crucial assistance in constructing solutions of the eigenvalue problem. 
Thus 
\eqn\pid{ 
\Delta^{(\vep)} \,\bigg (  {x\bx \over x-\bx}\bigg )^{\! 2\vep-1} = 
\bigg ( {x\bx \over  x-\bx}\bigg )^{\! 2\vep-1} \big ( \Delta^{(1-\vep)} 
+ 2 (1 - 2\vep ) \big ) \, , } 
so that letting 
\eqn\Fdu{
F^{(\vep)}{\!} _{\bl_1 \bl_2}(x,\bx) =  
\bigg ( {x\bx \over x-\bx}\bigg )^{\! 2\vep-1} \, 
{\tilde F}_{\bl_1  \bl_2}(x,\bx) \, , } 
gives equivalently 
\eqn\eigFD{ 
\Delta^{(-\vep+1)} {\tilde F}_{\bl_1 \bl_2}(x,\bx) = \big ( \bl_1 ( \bl_1 - 1 ) 
+ (\bl_2- 2\vep+1) ( \bl_2  - 2 ) \big ) \, 
{\tilde F}_{\bl_1 \bl_2}(x,\bx) \, . } 
When $\vep=1$ the conformal partial waves  are then reducible
to the $\vep =0$ case for which $ \Delta^{(\vep)}$ in \defD\ is just
a sum of two independent single variable differential operators.

Furthermore, with $v$ as in \defxz,
\eqn\Dveq{
\Delta^{(\vep)}(a,b) =
v^{-a-b} \Delta^{(\vep)}(-a,-b)\, v^{a+b} \, ,
}
which implies
\eqn\gveq{
F_{\lambda_1\lambda_2}(a,b;x,\bx) = 
v^{a+b} F_{\lambda_1\lambda_2}(-a,-b;x,\bx) \, ,
}
corresponding to combining \symF{a,b}.

\newsec{Integral Expressions}

Results for conformal partial waves can also be obtained in terms of
integral expressions for four point correlation functions. To describe
these results we first define for any conformal primary operator of spin
$\ell$, $\O^{(\ell)}_{\Delta,\mu_1 \dots \mu_\ell}$ an associated dual or
shadow operator by
\eqn\shad{
{\widetilde {\O}}^{(\ell)}_{d-\Delta,\mu_1 \dots \mu_\ell}(x) =
k_{\Delta,\ell} \, {1\over \pi^{{1\over 2}d}} \int \d^d y \;
{1\over \big ( (x-y)^2 \big ){}^{d-\Delta}} \, 
\I^{(\ell)}_{\mu_1 \dots \mu_\ell, \nu_1 \dots \nu_\ell}(x-y) \, 
\O^{(\ell)}_{\Delta,\mu_1 \dots \mu_\ell}(y) \, ,
}
where $\I^{(\ell)}_{\mu_1 \dots \mu_\ell,\nu_1 \dots \nu_\ell}(x)$ is the 
inversion
tensor for symmetric traceless tensors, formed from the symmetrised product
of $\ell$ inversion tensors $I_{\mu\nu}(x) = \delta_{\mu\nu} - 2 x_\mu x_\nu /x^2$.
The integral in \shad\ is divergent unless $\Delta < \half d$ but can be extended
to more general $\Delta$ by analytic continuation so that under conformal
transformations \shad\ defines a conformal primary operator of scale dimension 
$d-\Delta$. Choosing
\eqn\kD{
k_{\Delta,\ell} = {1\over (\Delta-1)_\ell} \, {\Gamma(d-\Delta+\ell)\over
\Gamma(\Delta - \half d)} \, ,
}
ensures that
\eqn\OOeq{
{\widetilde{\! \widetilde {\O}}}{}^{(\ell)}_{\Delta,\mu_1 \dots \mu_\ell}=
\O^{(\ell)}_{\Delta,\mu_1 \dots \mu_\ell} \, .
}
In general the shadow operator constructed in \shad\ does not belong
to the set of relatively local operators defining a unitary conformal
field theory but it plays a useful role in the present discussion.

To construct integral expressions we consider the conformally covariant
three point functions \Pet
\eqn\threep{\eqalign{
\big \langle \O^{(\ell)}_{\Delta,\mu_1 \dots \mu_\ell}(x) \, \vphi_3(x_3)\,
\vphi_4(x_4) \big \rangle = {}&  
{1\over (x_{34}{\!}^2 )^{{1\over 2}(\Delta_3+\Delta_4)}} \,
{\bar \F}^{(\ell)}_{\Delta,\mu_1 \dots \mu_\ell}(x;x_3,x_4) \, ,\cr
\big \langle  \vphi_1(x_1)\, \vphi_2(x_2) \, 
\O^{(\ell)}_{\Delta,\mu_1 \dots \mu_\ell}(x) \big \rangle = {}&
{1\over (x_{12}{\!}^2 )^{{1\over 2}(\Delta_1+\Delta_2)}} \,   
{\F}^{(\ell)}_{\Delta,\mu_1 \dots \mu_\ell}(x;x_1,x_2) \, ,
}
}
where
\eqn\FFeq{\eqalign{
{\bar \F}^{(\ell)}_{\Delta,\mu_1 \dots \mu_\ell}(x;x_3,x_4) = {}&
{\bar X}_{\{\mu_1} \dots {\bar X}_{\mu_\ell\}} \,
{(x_{34}{\!}^2 )^{\lambda_2} \over
\big ( (x-x_3)^2 \big ){}^{\lambda_2+b} \,
\big ( (x-x_4)^2 \big ){}^{\lambda_2- b} } \, , \cr
{\F}^{(\ell)}_{\Delta,\mu_1 \dots \mu_\ell}(x;x_1,x_2) = {}&
{X}_{\{\mu_1} \dots {X}_{\mu_\ell\}} \,
{(x_{12}{\!}^2 )^{\lambda_2} \over
\big ( (x-x_1)^2 \big ){}^{\lambda_2 - a} \,
\big ( (x-x_2)^2 \big ){}^{\lambda_2 + a}  \, , }
}
}
for
\eqn\XXeq{
{\bar X}_{\mu} = {(x_4 - x)_\mu \over (x_4-x)^2} -
{(x_3 - x)_\mu \over (x_3-x)^2}  \, , \qquad
{X}_{\mu} = {(x_1 - x)_\mu \over (x_1-x)^2} -
{(x_2 - x)_\mu \over (x_2-x)^2}  \, .
}
In \FFeq\ ${\bar X}_{\{\mu_1} \dots {\bar X}_{\mu_\ell\}}$, ${X}_{\{\mu_1} 
\dots {X}_{\mu_\ell\}}$ 
denotes symmetrisation and subtraction of traces. With the definition \shad\
and using integrals obtained in \DO
\eqn\threesh{\eqalign{
\big \langle {\widetilde \O}^{(\ell)}_{d-\Delta,\mu_1 \dots \mu_\ell}(x) \, 
\vphi_3(x_3)\, \vphi_4(x_4) \big \rangle = {}&  
{\gamma_{\blam_1,b} \over \gamma_{\lambda_1,b}} \, 
{1\over (x_{34}{\!}^2 )^{{1\over 2}(\Delta_3+\Delta_4)}} \, 
{\bar \F}^{(\ell)}_{d- \Delta,\mu_1 \dots \mu_\ell}(x;x_3,x_4) \, , \cr
\big \langle  \vphi_1(x_1)\, \vphi_2(x_2) \,
{\widetilde \O}{}^{(\ell)}_{d-\Delta,\mu_1 \dots \mu_\ell}(x) \big \rangle = {}&
{\gamma_{\blam_1,a} \over \gamma_{\lambda_1,a}} \, 
{1\over (x_{12}{\!}^2 )^{{1\over 2}(\Delta_1+\Delta_2)}} \,
{\F}^{(\ell)}_{d-\Delta,\mu_1 \dots \mu_\ell}(x;x_1,x_2) \, ,
}
}
for
\eqn\fda{
\gamma_{\lambda,a} = \Gamma\big (\lambda +a\big )
\, \Gamma\big (\lambda -a\big ) \, ,
}
and the abbreviations
\eqn\llb{
\blam_1 = \half (d - \Delta + \ell) \, , \quad
\blam_2 = \half (d - \Delta  - \ell) \, .
}

Assuming  \FFeq\ the expression for a four point function given by
\eqn\Fell{\eqalign{
F_\Delta^{(\ell)}(x_1,x_2,x_3,x_4) = {}&  
\gamma_{\lambda_1,a}^{\vphantom g} \gamma_{\blam_1,b} \, 
{1\over \pi^{{1\over 2}d}} \int \d^d x\; 
{\F}^{(\ell)}_{\Delta,\mu_1 \dots \mu_\ell}(x;x_1,x_2) \,
{\bar \F}^{(\ell)}_{d-\Delta,\mu_1 \dots \mu_\ell}(x;x_3,x_4) \, ,
}
}
is then conformally covariant so that 
\eqn\Msum{
{\textstyle \sum}_i M_{i\, AB} 
\Big ({1\over (x_{12}{\!}^2 )^{{1\over 2}(\Delta_1+\Delta_2)}\,
(x_{34}{\!}^2 )^{{1\over 2}(\Delta_3+\Delta_4)}} 
F_\Delta^{(\ell)}(x_1,x_2,x_3,x_4)\Big )  =0 \, .
 }
Since the three point functions in \threep\ have been constructed to be 
conformally covariant, so that
$(M_1+M_2+M)_{AB} \big \langle  \vphi_1(x_1)\, \vphi_2(x_2) \, 
\O^{(\ell)}_{\Delta,\mu_1 \dots \mu_\ell}(x) \big \rangle =0$,  we also have
\eqn\invF{
\big ( \half (M_1 + M_2)^{AB} (M_1+M_2)_{BA}  - C_{\Delta,\ell} \big ) \,
{1\over (x_{12}{\!}^2 )^{{1\over 2}(\Delta_1+\Delta_2)} } 
F_\Delta^{(\ell)}(x_1,x_2,x_3,x_4)=0 \, .
}
Using the relations between \threep\ and \threesh\ it is straightforward to 
see that the definition \Fell\ ensures that 
\eqn\shF{
F_\Delta^{(\ell)}(x_1,x_2,x_3,x_4) = F_{d-\Delta}^{(\ell)}(x_1,x_2,x_3,x_4) \, .
} 

The expression \Fell\ can be simplified by using
\eqn\XXb{
{X}_{\{\mu_1} \dots {X}_{\mu_\ell\}} \, {\bar X}_{\{\mu_1} \dots
{\bar X}_{\mu_\ell\}}
= {1 \over 2^\ell c_\ell} \,( X^2 {\bar X}^2 )^{{1\over 2}\ell}  \,
{\hat C}^{\,\vep}_\ell (t) \, , \qquad t =
{X \cdot {\bar X} \over (X^2 {\bar X}^2 )^{1\over 2}} \, ,
}
for ${\hat C}^{\,\vep}_\ell$ the modified Gegenbauer polynomial defined
in \CCl. Hence, as a consequence of conformal invariance,
\eqn\Fform{
F_\Delta^{(\ell)}(x_1,x_2,x_3,x_4) = { \gamma_{\lambda_1,a}^{\vphantom g}\gamma_{\blam_1,b} \over  2^\ell c_\ell} \, 
\bigg ( {x_{14}{\!}^2 \over x_{24}{\!}^2} \bigg)^{\! a} \!
\bigg ({ {x_{14}{\!}^2} \over x_{13}{\!}^2} \bigg)^{\! b}
f_\Delta^{(\ell)}(u,v) \, , 
}
where  
\eqnn\Fres
$$\eqalignno{
& {1\over \pi^{{1\over 2}d}} \int \d^d x\; {( X^2 {\bar X}^2 )^{{1\over 2}\ell}
\, {\hat C}^{\,\vep}_\ell (t) \over
\big ( (x-x_1)^2 \big ){}^{\lambda_2 - a} \,
\big ( (x-x_2)^2 \big ){}^{\lambda_2 + a} \,
\big ( (x-x_3)^2 \big ){}^{\blam_2+b} \,
\big ( (x-x_4)^2 \big ){}^{\blam_2- b} } \cr
&{}= {1\over (x_{12}{\!}^2 )^{\lambda_2}\,
(x_{34}{\!}^2 )^{\blam_2} }
\bigg ( {x_{14}{\!}^2 \over x_{24}{\!}^2} \bigg)^{\! a} \!
\bigg ({ {x_{14}{\!}^2} \over x_{13}{\!}^2} \bigg)^{\! b}
f_\Delta^{(\ell)}(u,v) \, .   &\Fres
}
$$
In \Fres\ we may write
\eqn\XXeq{
X^2 = {x_{12}{\!}^2 \over (x-x_1)^2  (x-x_2)^2} \, , \qquad
{\bar X}^2 = {x_{34}{\!}^2 \over (x-x_3)^2  (x-x_4)^2} \, ,
}
and
\eqn\XXbeq{\eqalign{
2 X \cdot {\bar X} ={}&    {x_{13}{\!}^2 \over (x-x_1)^2  (x-x_3)^2} - 
 {x_{23}{\!}^2 \over (x-x_2)^2  (x-x_3)^2}\cr
&{}  + {x_{24}{\!}^2 \over (x-x_2)^2  (x-x_4)^2} 
-  {x_{14}{\!}^2 \over (x-x_1)^2  (x-x_4)^2}  \, .
}
}

As a consequence of \invF
\eqn\DFeig{
\D \, f_\Delta^{(\ell)}(u,v) =
\half  C_{\Delta,\ell} \, f_\Delta^{(\ell)}(u,v) \, ,
}
so that, from \DGeig, $f_\Delta^{(\ell)}$ is a linear sum of the conformal 
partial waves $G_\Delta^{(\ell)}$ and $G_{d-\Delta}^{(\ell)}$.
The coefficients may be determined from the short distance limits 
$x_{12},x_{34}\to 0$. Assuming $\Delta < \half d$, so that $\lambda_2<\blam_2$, 
the leading contributions may be found directly using
\eqn\fllim{
{\F}^{(\ell)}_{\Delta,\mu_1 \dots \mu_\ell}(x;x_1,x_2) \limu{x_{12}\to 0}
(x_{12}{\!}^2 )^{\lambda_2} x_{12 \, \nu_1} \dots 
x_{12\, \nu_\ell} \, {1\over \big ( (x_1-x)^2 \big){}^\Delta}
\I^{(\ell)}_{\nu_1 \dots \nu_\ell, \mu_1\dots \mu_\ell}(x_1-x) \, ,
}
from \FFeq.
Hence
\eqn\Fllim{
\eqalign{
F_\Delta^{(\ell)}(x_1,x_2,x_3,x_4)  \limu{x_{12}\to 0} {}& 
{ {\gamma_{\lambda_1,a}\gamma_{\lambda_1,b}} \over k_{d-\Delta,\ell}}\,
(x_{12}{\!}^2 )^{\lambda_2} x_{12 \, \mu_1} \dots
x_{12\, \mu_\ell} \, 
{\bar \F}^{(\ell)}_{\Delta,\mu_1 \dots \mu_\ell}(x_1;x_3,x_4) \, .
}
}
while
\eqn\fbllim{
{\bar \F}^{(\ell)}_{\Delta,\mu_1 \dots \mu_\ell}(x_1;x_3,x_4) 
\limu{x_{34}\to 0} {1\over  ( x_{13}{\!}^2 )^\Delta} \,
\I^{(\ell)}_{\mu_1 \dots \mu_\ell, \nu_1\dots \nu_\ell}(x_{13}) \, 
x_{43 \, \nu_1} \dots x_{34\, \nu_\ell} \, 
(x_{43}{\!}^2 )^{\lambda_2} \, .
}
Since
\eqn\limv{
1-v  \limu{x_{12},x_{34}\to 0} 2 \, {1\over  x_{13}{\!}^2 }
x_{12\, \mu} I_{\mu\nu}(x_{13}) \, x_{43\, \nu} \, ,
}
then combining \Fllim\ and \fbllim\ in this limit
\eqn\xIx{
{1\over  (x_{13}{\!}^2)^\ell} \,x_{12 \, \mu_1} \dots x_{12\, \mu_\ell} \,
\I^{(\ell)}_{\mu_1 \dots \mu_\ell, \nu_1\dots \nu_\ell}(x_{13}) \, 
x_{43 \, \nu_1} \dots x_{43\, \nu_\ell} \sim
\big ( \half (1-v) \big )^\ell \, .
}
Comparing with the leading behaviour of $G_\Delta^{(\ell)}$ in \Gsim\ and using
\shF\ then gives
\eqn\fGG{
f_\Delta^{(\ell)}(u,v) =  {1\over k_{d-\Delta,\ell}}\, 
{\gamma_{\lambda_1,b} \over \gamma_{\blam_1,b}} \; G_\Delta^{(\ell)}(u,v) +
{1\over k_{\Delta,\ell}}\, {\gamma_{\blam_1,a} \over \gamma_{\lambda_1,a}} \;
G_{d-\Delta}^{(\ell)}(u,v)\, .
}

Explicit expressions can in principle be obtained by
expanding ${\hat C}^{\,\vep}_\ell (t)$ as a power series in $t$
and then using \XXbeq\ to expand $t^n$ leads  to a sum of elementary
conformal integrals \refs{\LR,\Mack}. Essentially equivalently,
we may use the recursion relation for Gegenbauer polynomials, 
generating all $ {\hat C}^{\,\vep}_\ell (t) $ starting from
$ {\hat C}^{\,\vep}_0 (t)=1$,  in the form
\eqn\refC{
(\ell + 2\vep ) \, {\hat C}^{\,\vep}_{\ell+1} (t) = 2(\ell +\vep) \, t \, 
{\hat C}^{\,\vep}_{\ell} (t) - \ell \, {\hat C}^{\,\vep}_{\ell-1} (t) \, ,
}
which in \Fres\ leads to
\eqnn\recurF$$\eqalignno{
(\ell + d - 2)&  \, f_\Delta^{(\ell+1)} (a,b;u,v) \cr
= {}& (\ell + \half d -1) \, u^{- {1\over 2}}  \big ( 
f_\Delta^{(\ell)} (a-\half,b+\half ;u,v) - 
v f_\Delta^{(\ell)} (a+\half,b+\half ;u,v) \cr
\noalign{\vskip -1pt}
&\hskip 2.5cm + f_\Delta^{(\ell)} (a+\half,b-\half ;u,v)
- f_\Delta^{(\ell)} (a-\half,b-\half ;u,v) \big ) \cr
& \ \ {}- \ell\,  f_\Delta^{(\ell-1)} (a,b;u,v) \, . & \recurF
}
$$
For $d=2,4$ this recurrence relation
was solved in \DO\ by using corresponding relations for the
single variable $g_p$ functions.

For $\ell=0$ the conformal integral in \Fres\ may be evaluated as a  
Mellin-Barnes transform \Syman
\eqn\MBf{\eqalign{
f_\Delta^{(0)}(u,v) = {1\over \gamma_{\lambda,a} \gamma_{\blam,b}}\,
\int_{C_{s,t}}\!\!\!\!  \d \mu_{s,t} \; 
\Gamma(-t)\, \Gamma(-t-a-b) \, & \Gamma(-r+a) \, \Gamma(-r+b)  \cr
\noalign{\vskip -6pt}
& {} \times \, \Gamma(\lambda - s )\,  \Gamma(\blam -s) \, u^s v^t \, ,
}
}
for $\lambda=\half\Delta, \blam = \half(d-\Delta)$ and 
\eqn\rst{
\d \mu_{s,t} = {1\over (2\pi i)^2} \, \d s \, \d t \, , \qquad r+t + s =0 \, .
}
The integration contours $C_{s,t}$ for $s,t$ run from $-i \infty$ to 
$ i \infty$ such that
\eqn\contour{
{\sl Re} \, s  < \lambda, \blam \, , \qquad {\sl Re} \, t < 0, -a-b \, , \qquad
{\sl Re} \, r  <  a,b \, . } 
If the $s$ contour is closed to the right the poles at $s=\lambda+n$ and 
$s=\blam +n $, for $n=0,1,2,\dots$, generate the conformal partial wave 
$G_\Delta^{(0)}$ and its dual $G_{d-\Delta}^{(0)}$ in accordance
with \fGG. In the remaining $t$-integration  it is useful to note that, 
for $-a-s,-b-s < c < 0 , -a-b$,
\eqn\tint{\eqalign{
{}& {1\over 2\pi i}
\int_{c -i \infty}^{c+ i \infty} \!\!\!\!\!\!\!\! \d t \;
\Gamma(-t)\Gamma(-t-a-b) \, \Gamma(s+t+a) \, \Gamma(s+t+b) \, v^t \cr
&{} = {1\over \Gamma(2s)} \, \gamma_{s,a} \gamma_{s,b} \, 
F(s+a,s+b;2s;1-v) \, .
}
}
By restricting the contours $C_{s,t}$ so as to pick up only the 
poles at $s=\lambda+n$ then \MBf\  gives
for the $\ell=0$ conformal partial wave a double power series
\eqn\Gzero{ 
G_{2\lambda}^{(0)}(u,v) = \sum_{n,m\ge 0} 
{(\lambda-a)_n(\lambda-b)_n \over n! \,(2\lambda - \vep)_n} \, 
{(\lambda+a)_{n+m}(\lambda+b)_{n+m} \over m! \,(2\lambda)_{2n+m}} 
\; u^{\lambda+n} (1-v)^m \, ,
} 
where convergence requires for the $m$-sum $v<1$ and for large $n={\rm O}(m)$
\eqn\conv{
(1-v+u)^2 < 4u \, .
}
Remarkably this coincides with the allowed region of $u,v$ for
$x_i$ vectors belonging to $d$-dimensional Euclidean space
(letting $x_1=0$, $x_4\to \infty$ then $1-v+u = 2\, x_2 \cdot x_3 /x_3{\!}^2$
and $u= x_2{\!}^2/x_3{\!}^2$ so that \conv\ is equivalent to the usual
Schwarz inequality). With \defxz\ \conv\ becomes $(x-\bx)^2 <0$.

For general $\ell$ \MBf\ may be extended to
\eqn\MBfl{\eqalign{
f_\Delta^{(\ell)}(u,v)
 = {}& {1\over \gamma_{\lambda_1,a} \gamma_{\blam_1,b}}\,
\int_{C_{s,t}}\!\!\!\!  \d \mu_{s,t} \;
\Gamma(-t)\, \Gamma(-t-a-b) \, \Gamma(-r+a) \, \Gamma(-r+b) \cr
\noalign{\vskip -6pt}
& \hskip 3cm {} \times \,
\Gamma(\lambda_2 - s )\,  \Gamma(\blam_2 -s) \, \alpha_\ell(a,b;r,t,s) \, 
u^s v^t \, , \cr
}
}
so that from \MBf\ $\alpha_0=1$.
Since the symmetry \symG\ also applies to $f_\Delta^{(\ell)}(u,v)$ we must have
\eqn\symal{
\alpha_\ell(a,b;r,t,s) = (-1)^\ell \alpha_\ell(-a,b;t+b,r-b,s)
= (-1)^\ell \alpha_\ell(a,-b;t+a,r-a,s)  \, .
}
Applying the eigenvalue equation \DFeig, using the result \LG\ 
for the operator $\D$ and suppressing the $a,b$ arguments in $\alpha_\ell$, gives
\eqn\recal{\eqalign{
& t (t+a+b) \big ( \alpha_\ell(r+1,t-1,s) - \alpha_\ell(r,t,s) \big )\cr
& {}+ (r-a)(r-b) \big ( \alpha_\ell(r-1,t+1,s) - \alpha_\ell(r,t,s) \big )\cr
&{} - (s-\lambda_2)(s - \blam_2) 
\big ( \alpha_\ell(r+1,t,s-1) + \alpha_\ell(r,t+1,s-1) 
- 2\, \alpha_\ell(r,t,s) \big )\cr
&{} =\big (  2s ( \half d - \lambda_2 - \blam_2 ) + 2 \lambda_2 \blam_2
+ \half C_{\Delta,\ell} \big ) \, \alpha_\ell(r,t,s)  
= \ell (2s+ \ell-1) \, \alpha_\ell(r,t,s) \, .
}
}
using \Ceig\ for $C_{\Delta,\ell}$ and  \lll,\llb\ for $\lambda_2,\blam_2$.
Equivalent difference equations have been given in \pene\ in a more general
context. In order that the analytic structure of the conformal partial
waves remain valid for any $\ell$ we require that $\alpha_\ell$ is a 
polynomial in $r,t,s$. For any $\ell=0,1,2 \dots$ there are solutions  
$\alpha_\ell(r,t,s) = \sum_{n+m\le \ell} c_{nm}  \, w^n s^m$ 
where $w=t-r + a +b$ which depend on the $\half (\ell+1)(\ell+2)$
coefficients $c_{nm}$. With $w$ as defined above \symal\ requires 
that these are constrained by
$c_{nm}(a,b) = (-1)^{\ell-n} c_{nm}(-a,b) = (-1)^{\ell-n} c_{nm}(a,-b) $.
For the assumed expression for $\alpha_\ell$
the recurrence relation $\recal$  generates terms $w^n s^m$
with $n+m\le \ell+1$ but the equation is an identity when $n+m = \ell+1$
and also $n = \ell , m=0$ so that there remain $\half \ell (\ell+3)$
linearly independent equations determining $c_{nm}$ uniquely up to an
overall constant factor. The existence of polynomial solutions is
crucially dependent on the factor $\ell(2s+\ell-1)$ appearing on the
right hand side of \recal. This in turn is dependent on the choice of
$\lambda_2,\blam_2$ in \MBfl. 
As will be demonstrated below these polynomial solutions ensure
that the required boundary conditions \Gsim\ are satisfied.

More directly from \recurF
\eqnn\recurA
$$\eqalignno{
& (\ell + d -3) \, \alpha_\ell (a,b;r,t,s) \cr
&{}= (\ell + \half d -2)  \big (
(\lambda_1+a-1)(\blam_1-b-1) \, (b-r) \,
\alpha_{\ell-1} (a-\half,b+\half ;r-\half,t,s+\half) \cr
\noalign{\vskip -2pt}
&\hskip 2.3cm {}+
(\lambda_1-a-1)(\blam_1-b-1) \, t \,
\alpha_{\ell-1} (a+\half,b+\half ;r+\half,t-1,s+\half) \cr
\noalign{\vskip -2pt}
&\hskip 2.3cm +
(\lambda_1-a-1)(\blam_1+b-1) \, (a-r) \, 
\alpha_{\ell-1} (a+\half,b-\half ; r-\half , t, s+\half ) \cr
\noalign{\vskip -2pt}
&\hskip 2.3cm +
(\lambda_1+a-1)(\blam_1+b-1) \, (t+a+b)\, 
\alpha_{\ell-1} (a-\half,b-\half ;r-\half , t , s+\half ) \big ) \cr
& \quad {}- (\ell-1)\, 
(\lambda_1+a-1)(\lambda_1-a-1)(\blam_1+b-1)(\blam_1-b-1)\,
(\lambda_2-s)(\blam_2-s) \cr
\noalign{\vskip -2pt}
&\hskip 9cm {} \times \alpha_{\ell-2} (a,b;r,t,s) \, , & \recurA
}
$$
where the $\lambda_1, \blam_1$ factors are a consequence of the
normalisation of the integral in \MBfl. Starting $\alpha_0=1$ this 
gives $(d-1)_{\ell-1} \, \alpha_\ell$ iteratively, for any $\ell=1,2, \dots$, 
as a polynomial in all variables.

For $\ell=1$
\eqn\alpo{
\alpha_1(r,t,s) =  \lambda_2 \blam_2 \, w - ab \, ( s - \half d + 1 ) \, ,
}
and for $\ell=2,a=b=0$,
\eqn\alto{\eqalign{ 
{2(d-1) \over (\lambda_2+1)(\blam_2+1)} \, \alpha_2(r,t,s) = {}&
(2\lambda_2+1)(2\blam_2+1)(\lambda_2 + \blam_2 + 2) \, w^2 \cr
\noalign{\vskip -7pt} 
&{} - ( \lambda_2 + \blam_2 +  2 \lambda_2 \blam_2) \, 
s ( s - 2\lambda_2 -2 \blam_2 - 2) \cr
\noalign{\vskip -1pt} 
&{} - 2 \lambda_2 \blam_2 (\lambda_2+1)(\blam_2+1) \, .
}
}

For any $\ell$ in order to satisfy the boundary condition \Gusim\ it is
sufficient to consider only the contribution of the pole at $s=\lambda_2$ 
and then require
\eqn\speca{
\alpha_\ell(r,t,\lambda_2) = (d-\Delta - 1)_\ell c_\ell \; \beta_\ell(r,t) \, ,
}
where 
\eqn\defbe{\eqalign{
\beta_\ell(r,t)= {}&
\sum_{n=0}^\ell {\ell \choose n} (-1)^n \,
(-t)_n (-t-a-b)_n (-r+a)_{\ell -n} (-r+b)_{\ell-n} \cr
= {}& (a-r)_\ell (b-r)_\ell \,
{}_3 \! F_2 \Big ( {-\ell,-t,-t-a-b \atop r-a - \ell  +1, r-b - \ell  +1}; 1
\Big ) \, ,
}
}
since then, using \tint, the $t$-integral  gives
$\sum_{n=0}^\ell {\ell \choose n} (-1)^n  v^n \, 
F (\lambda_1+a,\lambda_1+b; 2 \lambda_1;1-v)$, so that
there is the necessary $(1-v)^\ell$ factor required  by \Gusim. In 
writing \defbe\ in terms of a ${}_3 \! F_2$-function it is useful to note the
Pochhammer symbol identity
\eqn\Poch{
(x)_{N-n} = (-1)^n \, {(x)_N \over (1-x-N)_n} \, .
}

Defining also
\eqn\hypert{\eqalign{
\gamma_\ell(r,t)  = {}& (b-r)_\ell ( \lambda_2 + a )_\ell \,
{}_3 \! F_2 \Big ( {-\ell,-t,- \lambda_1  +a+1   \atop r- b -\ell  +1,
\lambda_2 + a}; 1 \Big ) \, ,\cr
\delta_\ell(r,t)  = {}& (-1)^\ell (-t-a-b)_\ell ( \lambda_2 + a )_\ell \,
{}_3 \! F_2 \Big ( {-\ell,-r+a,- \lambda_1  +a+1   \atop t+ a+b -\ell  +1,
\lambda_2 + a}; 1 \Big ) \, ,
}
}
then, with the aid of Sheppard identity \Spec\foot{
$$\eqalign{
\noalign{\vskip -8pt}
(\gamma)_\ell(\delta)_\ell\; {}_3 \! F_2(-\ell,\alpha,\beta;\gamma,\delta;1) = {}&
(\gamma)_\ell(\delta-\alpha)_\ell 
\; {}_3 \! F_2 \Big ({-\ell, \alpha, \delta-\beta \atop \gamma, 
\alpha - \delta + 1 - \ell};1 \Big ) \cr
= {}& (\gamma-\alpha)_\ell (\delta-\alpha)_\ell
\; {}_3 \! F_2 \Big ({-\ell, \alpha, \alpha+\beta-\gamma-\delta+1-\ell \atop 
\alpha - \gamma+ 1 -\ell, \alpha - \delta + 1 - \ell};1 \Big ) \, .}
$$},
\eqn\relabc{
\beta_\ell(r,t) = \gamma_\ell(r,t) = \delta_\ell(r,t) \qquad \hbox{for} \quad 
- r -t = \lambda_2 \, ,
}
and 
\eqn\recurbc{\eqalign{
\gamma_\ell(a,b;r,t) ={}& (\lambda_1+a-1)(b-r) \, 
\gamma_{\ell-1}(a-\half,b+\half; r-\half,t) \cr
\noalign{\vskip -2pt}
&\quad {} + (\lambda_1-a-1)t \, \gamma_{\ell-1}(a+\half,b+\half;
r+\half,t-1) \, , \cr
\delta_\ell(a,b;r,t) ={}& (\lambda_1+a-1)(t+a+b) \, 
\delta_{\ell-1}(a-\half,b-\half; r-\half,t) \cr
\noalign{\vskip -2pt} 
&\quad {} + (\lambda_1-a-1)(a-r) \, \delta_{\ell-1}(a+\half,b-\half;
r-\half,t) \, , \cr
}
}
which ensures that \speca\ satisfies \recurA\ for $s=\lambda_2$. In
the above $\beta_\ell$ satisfies both symmetries in \symal\ while 
$\gamma_\ell(a,b;r,t) = (-1)^\ell \gamma_\ell(-a,b;t+b,r-b)$, and similarly for
$\delta_\ell$, but $ (-1)^\ell \gamma_\ell (a,-b;t+a,r-a)=
\delta_\ell(a,b;r,t)$.

If $\blam_1 = 1 \mp b$, or $\lambda_2 = \vep \pm b$, then
\recurA\ simplifies to three term relations which have the solutions
\eqn\twA{
\alpha_\ell(r,t,s)\big |_{\blam_1 = 1-b} = (-1)^\ell (2b)_\ell c_\ell \,
\gamma_\ell(r,t) \, , \qquad
 \alpha_\ell(r,t,s)\big |_{\blam_1 = 1+b} = (2b)_\ell c_\ell \,
\delta_\ell(r,t) \, ,
} 
There are analogous solutions
of \recurA\ for $\lambda_1 = 1 \pm a$ but, given
\fGG\ and \MBfl, these are irrelevant for the conformal 
partial wave $G_\Delta^{(\ell)}$. The solutions for $\alpha_\ell$
in \twA\ in terms of  ${}_3\!F_2$ functions correspond to the 
leading twist conformal partial waves.

In general $\alpha_\ell$ has no very compact form but it may be expressed
as a quadnomial sum
\eqn\alres{\eqalign{
(d-2)_\ell \, \alpha_\ell = {}& \sum_{m+n+p+q=\ell} 
{\ell! \over m!\, n! \,p! \, q!} \,
(-1)^{p+n} (2\blam_2+\ell-1)_{\ell-q} (2\lambda_2+\ell-1)_{n} \cr
\noalign{\vskip -3pt}
&\quad {}\times (\blam_1+a-q)_q(\blam_1+b-q)_q (\lambda_1+a-m)_m
(\lambda_1 +b-m)_m \cr
\noalign{\vskip -2pt}
&\quad {}\times (d-2+\ell+n-q)_q (\half d-1)_{\ell-q} (\half d-1+a+b + n)_p\cr
\noalign{\vskip -2pt}
&\qquad {}\times (\lambda_2 - s)_{p+q} \, (-t)_n \, ,
}
}
where the various Pochhammer symbols may be rewritten using \Poch\ and
\eqn\Pch{
(x-n)_n = (-1)^n \, (1-x)_n \, .
}
The result \alres\ is symmetric under $a\leftrightarrow b$. It can be reexpressed 
in a variety of similar forms, such as those obtained by imposing 
\symal\ or letting $\lambda_2, \lambda_1 \leftrightarrow \blam_2,\blam_1$, 
albeit with no significant simplification. Such identities may be obtained
from \alres\ by repeatedly writing some of the summations in terms of a 
terminating ${}_3{\!}F_2(1)$ function and using the Sheppard identity.

In two dimensions there are significant simplifications using complex,
or light cone, coordinates. If $x^2 = z \bz, \, x'^2 = z' \bz'$ then
$(x^2 x'^2)^{{1\over 2}\ell} {\hat C}_\ell^{\, 0}\big ( x \cdot x'/
(x^2 x'^2)^{{1\over 2}} \big )
= \half \big ( ( z \bz')^\ell + ( \bz z')^\ell \big )$ and hence
\eqn\twof{\eqalign{
& {( X^2 {\bar X}^2 )^{{1\over 2}\ell}  \, {\hat C}^{\, 0}_\ell (t) \over
\big ( (x-x_1)^2 \big ){}^{\lambda_2 - a} \,
\big ( (x-x_2)^2 \big ){}^{\lambda_2 + a} \,
\big ( (x-x_3)^2 \big ){}^{\blam_2+b} \,
\big ( (x-x_4)^2 \big ){}^{\blam_2- b} } \cr
& \quad {}= \half ( z_{12} \, \bz_{43} )^\ell \, f_{\lambda_1}(z;z_i) \, 
f_{\lambda_2}(\bz;\bz_i) +  \half ( \bz_{12} \, z_{43} )^\ell
f_{\lambda_2}(z;z_i) \, f_{\lambda_1}(\bz;\bz_i) \, ,
}
}
where
\eqn\ffour{
f_{\lambda}(z;z_i)  = { 1\over (z - z_1)^{\lambda-a} \, 
 (z- z_2)^{\lambda+ a} \, (z- z_3)^{1-\lambda+ b} \,  (z- z_4)^{1-\lambda- b}}\, .
}
The essential integral, which is evaluated in appendix A, has the form
\eqn\ffint{\eqalign{
{1\over \pi} \int  & \d^2 x \; ( z_{12} \, \bz_{43} )^\ell \, f_{\lambda_1}(z;z_i) \, 
f_{\lambda_2}(\bz;\bz_i) \cr
={}&   {1\over (x_{12}{\!}^2 )^{\lambda_2}\, 
(x_{34}{\!}^2 )^{1-\lambda_1} }
\bigg ( {x_{14}{\!}^2 \over x_{24}{\!}^2} \bigg)^{\! a} \!  
\bigg ({ {x_{14}{\!}^2} \over x_{13}{\!}^2} \bigg)^{\! b}  \cr
&{}\times \bigg ( {\gamma_{\lambda_1,b}\over \gamma_{1-\lambda_2,b}}  \,
{\Gamma(1-2\lambda_1)\over \Gamma(2\lambda_2)} \,
g_{\lambda_1}(x) \, g_{\lambda_2}(\bx) + 
{\gamma_{1-\lambda_1,a}\over \gamma_{\lambda_2,a}}  \,
{\Gamma(2\lambda_1-1)\over \Gamma(2-2\lambda_2)} \,
g_{1-\lambda_1}(x) \, g_{1-\lambda_2}(\bx) \bigg ) \, ,
}
}
for
\eqn\xzz{
x= {z_{12}\, z_{34} \over z_{13} \, z_{24}} \, , \qquad
\bx= {\bz_{12}\, \bz_{34} \over \bz_{13} \, \bz_{24}} \, .
}
Since $1-x= z_{14}\, z_{23}/z_{13}\, z_{24}$ it is easy to see that
this is in accord with the definition \defuv\ of $x,\bx$. 
Using repeatedly $\Gamma(x)\Gamma(1-x) = \pi/\sin \pi x$, from
which follow 
the identities ${\Gamma(1-2\lambda_1)/ \Gamma(2\lambda_2)}
= {\Gamma(1-2\lambda_2)/ \Gamma(2\lambda_1)}$ and
${\gamma_{1-\lambda_1,a}/ \gamma_{\lambda_2,a}} 
= {\gamma_{1-\lambda_2,a}/\gamma_{\lambda_1,a}}  $, then
from \Fres, \twof\ and \fGG\ in two dimensions we get
\eqn\Gtwo{
G_{\Delta}^{(\ell)}(u,v)  = \half (
g_{\lambda_1}(x) \, g_{\lambda_2}(\bx) + g_{\lambda_1}(\bx) \, 
g_{\lambda_2}(x) \big ) \, .
}

\newsec{Recurrence Relations}

In the discussion of orthogonal polynomials recurrence relations involving
two or more terms play a significant role. Here we derive analogous
results for conformal partial waves, extending and simplifying the results
given in \DT. They play a role in combining conformal partial waves into
superconformal blocks.

We first consider recurrence relations for the single variable functions
$g_p$ satisfying \defg\ and whose derivation is a model for obtaining
recurrence relations for the two variable conformal partial waves. 
With the differential operator $D_x(a,b)$ defined by \hyper\ it is 
straightforward to obtain
\eqn\Dabh{\eqalign{
D_x(a+\half,b+\half) \, x^{-{1\over 2}} = {}& x^{-{1\over 2}} \, D_x(a,b)
-  x^{{1\over 2}} {\d \over \d x}+ {\ts {3\over 4}} \, x^{-{1\over 2}} \, , \cr
D_x(a+\half,b+\half) \, x^{{1\over 2}} {\d \over \d x}  = {}& 
\Big ( x^{{1\over 2}} {\d \over \d x} - x^{-{1\over 2}} \Big ) \, D_x(a,b)
- \quar \, x^{{1\over 2}} {\d \over \d x} \, .
}}
Since $g_p$ is defined by being a solution of \defg,
\Dabh\ ensures that $x^{-{1\over 2}} \, g_p(a,b;x)$ and 
$x^{{1\over 2}} {\d \over \d x} \,  g_p(a,b;x)$
can be expressed in terms of $ g_{p'}(a+\half, b+\half;x)$ for just
$p'= p \pm \half$ in the form
\eqn\ggeq{\eqalign{
x^{-{1\over 2}} \, g_p(a,b;x) = {}& g_{p-{1\over 2}}(a+\half, b+\half;x)
- \sigma_p \, g_{p+ {1\over 2}}(a+\half, b+\half;x) \, , \cr
x^{{1\over 2}} {\d \over \d x}\, g_p(a,b;x) = {}& p\, 
g_{p-{1\over 2}}(a+\half, b+\half;x)
+(p-1) \sigma_p \,  g_{p+ {1\over 2}}(a+\half, b+\half;x) \, .
}}
In the result for $x^{-{1\over 2}} \, g_p(a,b;x)$ the coefficient of
$g_{p-{1\over 2}}(a-\half, b+\half;x)$ is determined by requiring the
normalisation condition \glim\ for the leading $x^{p-{1\over 2}}$ term.
By applying \gab\ and writing in \ggeq\ $\sigma_p = \sigma_p(a,b)$ then
\ggeq\ may be extended to
\eqn\threeg{\eqalign{
x^{-{1\over 2}} \, g_p(a,b;x) = {}& g_{p-{1\over 2}}(a\pm\half, b\mp\half;x)
+ \sigma_p(\pm a,\mp b) \, g_{p+ {1\over 2}}(a\pm\half, b\mp\half;x) \, , \cr
x^{-{1\over 2}} \, (1-x)g_p(a,b;x) = {}& g_{p-{1\over 2}}(a-\half, b-\half;x)
- \sigma_p(-a,-b) \, g_{p+ {1\over 2}}(a-\half, b-\half;x) \, . \cr
}}
Assuming the normalisation condition \glim\ for $g_{p + {1\over2}}$, $\sigma_p$ 
may be determined to be
\eqn\sigp{
\sigma_p(a,b) = {(p+a)(p+b)\over 2p(2p-1)} \, .
}

There are also relations between $g_p$ for differing $p$ but the same $a,b$.
Although they follow from standard identities a  derivation 
which may be extended later to conformal partial waves is obtained by 
first defining
\eqn\defff{
f_0 = {1\over x} - \half   \, , \qquad 
f_1 = (1-x) {\d \over \d x} -\half (a+b) \, ,
}
which satisfy the commutation relations 
\eqn\Dff{
\big [ D_x , f_0 \big ] = -2 f_1 + 2 f_0 \, , \qquad
\big [ D_x , f_1 \big ] = - 2 f_0 \, D_x - ab \, .
}
These relations ensure that $f_0 g_p, \, f_1 g_p$ may be expanded in terms
of only $g_{p\pm 1}, \, g_p$ giving the four term relations 
\eqna\idgp
$$\eqalignno{
f_0  \, g_p = {}& g_{p-1} + 
\alpha_p \, g_p + \beta_p \, g_{p+1} \, , &\idgp{a} \cr
f_1 \,  g_p = {}& p \, g_{p-1} + \alpha_p \, g_p -(p-1) \beta_p \, g_{p+1} \, .
& \idgp{b}
}
$$
where the coefficient of $g_{p-1}$ is determined by imposing \glim\ and
\Dff\ determines
\eqn\idab{ 
\alpha_p = - {ab \over 2p(p-1)} \, .
}
To determine $\beta_p$ it is necessary to take account of the normalisation
of $g_{p+1}$. This may be achieved by using
$[f_1 , f_0 ] = - f_0{\!}^2 + \quar$ from which we may obtain
\eqn\idb{
\beta_p = {(p+a)(p+b)(p-a)(p-b) \over 4p^2(2p-1)(2p+1)} \, .
}
By iterating \threeg\ $\beta_p = \sigma_p(-a,b)\, \sigma_{p+{1\over 2}}
(a-\half,-b-\half)$.

The single variable recurrence relations for $g_p$ can be extended
to $F_{\lambda_1\lambda_2}$ in a similar fashion. The relations satisfy
the symmetry
\eqn\symFF{
F^{(\vep)}{\!}_{\lambda_2 - \vep  \, \lambda_1+ \vep}(x,\bx)
= F^{(\vep)}{\!}_{\lambda_1 \lambda_2}(x,\bx) \, .
}
This is in part a consequence of $C_{\Delta,\ell} = C_{\Delta,-\ell-d+2}$,
$D_{\Delta,\ell} = D_{\Delta,-\ell-d+ 2}$.
The symmetry relation \symFF\ depends on the choice of normalisation in \cell\
(with the normalisation in \CCl\ we may identify ${\hat C}^{\,\vep}_\ell(t)
= {\hat C}^{\, \vep}_{-\ell- 2 \vep}(t)$).

We first consider the corresponding operators acting on two variable 
conformal partial waves to those considered in \Dabh. Defining
\eqn\HHH{\eqalign{
\H_1 = {}& (x\bx)^{-{1\over 2}} \bigg ( x {\pr \over \pr x}
 + \bx {\pr \over \pr \bx} \bigg )  \, , \quad
\H_2 = (x\bx)^{1\over 2} \bigg ( {\pr^2 \over \pr x \pr \bx} - 
\vep \, {1\over x- \bx} \Big ( {\pr \over \pr x} - {\pr \over \pr \bx} \Big )
\bigg ) \, , \cr
\H_3 = {}& (x\bx)^{-{1\over 2}} 
\bigg (x {\pr \over \pr x}-\bx {\pr \over \pr \bx}
+ 2 \vep \, \, {x+\bx \over x- \bx} \bigg )
\big ( D_x(a,b) - D_\bx(a,b) \big ) \, ,
}
}
then these satisfy the algebra
\eqn\HHHD{\eqalign{
\Delta^{(\vep)}(a+\half,b+\half) \, (x\bx)^{-{1\over 2}} = {}&
(x\bx)^{-{1\over 2}} \Delta^{(\vep)}(a,b) + ( {\ts{3\over 2}} + \vep)
(x\bx)^{-{1\over 2}} - \H_1 \, , \cr
\Delta^{(\vep)}(a+\half,b+\half) \, \H_1 = {}&
\big ( \H_1 - (x\bx)^{-{1\over 2}} \big )  \Delta^{(\vep)}(a,b)  + 
( {\ts{1\over 2}} + \vep) \H_1  - 2 \H_2 \, , \cr
\Delta^{(\vep)}(a+\half,b+\half) \, \H_2 = {}&
\big ( \H_2 - \half \H_1 - \vep (x\bx)^{-{1\over 2}} \big )  
\Delta^{(\vep)}(a,b) - ({\ts{1\over 2}} + \vep) \H_2  +\half  \H_3 \, ,
}
}
and
\eqn\HHC{
\Delta^{(\vep)}(a+\half,b+\half) \, \H_3 = \H_3  \, \Delta^{(\vep)}(a,b)  + 
( {\ts{1\over 2}} - \vep) \H_3  -  (x\bx)^{-{1\over 2}} 
\Delta_4{\!}^{(\vep)}(a,b) \, .
}
Here $\Delta_4$ is the fourth order  Casimir operator
\eqn\Casz{
\Delta_4{\!}^{(\vep)} (a,b)
= \bigg ( {x\bx \over x-\bx} \bigg )^{\! 2 \vep} 
\big ( D_x(a,b) - D_\bx(a,b) \big ) \,
\bigg ( {x\bx \over x-\bx} \bigg )^{\! - 2\vep}
\big ( D_x(a,b) - D_\bx(a,b) \big ) \, ,
}
which satisfies
\eqn\DDcom{
\big [ \Delta^{(\vep)}(a,b) , \Delta_4{\!}^{(\vep)}(a,b) \big ] = 0 \, .
}
Acting on conformal partial waves
\eqn\eigC{
\Delta_4{\!}^{(\vep)}  F_{\lambda_1\lambda_2}= c_{4,\lambda_1\lambda_2} \,
F_{\lambda_1\lambda_2} \, ,
}
where
\eqn\eigDf{\eqalign{
c_{4,\lambda_1\lambda_2} = {}& (\lambda_1-\lambda_2)(\lambda_1-\lambda_2+2\vep)
(\lambda_1+\lambda_2-1)(\lambda_1+\lambda_2-1-2\vep) \, .
}
}
The formula for $c_{4,\lambda_1\lambda_2}$ may be obtained 
by considering the limit $\bx \to 0$ in \eigC\ when 
$\Delta_4{\!}^{(\vep)}(a,b) \sim  D_x(a,b) - \bx^2 {\pr^2 \over \pr \bx^2}
+ 4 \vep \,  \bx {\pr \over \pr \bx} -2\vep (1+2\vep)$ and
using  \Fxsim. The expression \eigDf\ is related to the quartic Casimir given 
in \CasO\ since, with the
relations \lll, $D_{\Delta,\ell} = -2  c_{4,\lambda_1\lambda_2} + 4
c_{\lambda_1\lambda_2}{\!\!}^2 + d(d-1)\, c_{\lambda_1\lambda_2}$.

In general $(x\bx)^{-{1\over 2}} F_{\lambda_1\lambda_2}(a,b;x,\bx)$
has an expansion in terms of the linearly independent basis
$F_{\lambda_1-{1\over 2}+n\, \lambda_2-{1\over 2}+m}(a+\half,b+\half;x,\bx)$
for $n,m=0,1,2,\dots$. However  consistency conditions arising from \HHHD\ and \HHC\
restrict this to only four terms of the form
\eqn\Frec{\eqalign{
 (x\bx)^{-{1\over 2}} F_{\lambda_1\lambda_2}(a,b) = {}&
r\, F_{\lambda_1-{1\over 2}\, \lambda_2-{1\over 2}}(a+\half,b+\half) +
s\, F_{\lambda_1+{1\over 2}\, \lambda_2-{1\over 2}}(a+\half,b+\half) \cr
&{}+ t\, F_{\lambda_1-{1\over 2}\, \lambda_2+{1\over 2}}(a+\half,b+\half) +
u\, F_{\lambda_1+{1\over 2}\, \lambda_2+{1\over 2}}(a+\half,b+\half) \, .
}}
Using \HHHD\ we then get from \Frec\ for $i=1,2,3$
\eqn\FFrec{\eqalign{
\H_i F_{\lambda_1\lambda_2}(a,b) = {}&
r_i \, F_{\lambda_1-{1\over 2}\, \lambda_2-{1\over 2}}(a+\half,b+\half) +
s_i\, F_{\lambda_1+{1\over 2}\, \lambda_2-{1\over 2}}(a+\half,b+\half) \cr
&{}+ t_i\, F_{\lambda_1-{1\over 2}\, \lambda_2+{1\over 2}}(a+\half,b+\half) +
u_i\, F_{\lambda_1+{1\over 2}\, \lambda_2+{1\over 2}}(a+\half,b+\half) \, ,
}}
where
\eqn\rstu{\eqalign{
r_1 = {}& (\lambda_1+\lambda_2)r \, , \quad 
s_1 = -(\lambda_1-\lambda_2-1)s \, ,\quad
t_1 = (\lambda_1-\lambda_2 + 1 +  2\vep)t \, , \cr
& \quad
u_1 = - (\lambda_1+\lambda_2- 2- 2\vep) u \, , \cr
r_2 = {}& (\lambda_1 + \vep ) \lambda_2 r \, , \quad  
s_2 = - (\lambda_1 -1- \vep) \lambda_2 s\, , \quad 
t_2 = - (\lambda_1 + \vep ) (\lambda_2 - 1 - 2\vep)t\, ,\cr 
& \quad u_2 = (\lambda_1 -1- \vep ) (\lambda_2 - 1 - 2\vep)u \, , \cr
& \quad r_3 =  (\lambda_1-\lambda_2) 
(\lambda_1-\lambda_2+ 2\vep)(\lambda_1+\lambda_2-1) r \, , \cr
& \quad s_3 = - (\lambda_1-\lambda_2) 
(\lambda_1+\lambda_2 - 1)(\lambda_1+\lambda_2-1- 2\vep)s \, , \cr
& \quad t_3 = (\lambda_1-\lambda_2 + 2 \vep)
(\lambda_1+\lambda_2 - 1)(\lambda_1+\lambda_2-1- 2\vep)s \, , \cr
& \quad u_3  = - (\lambda_1-\lambda_2)
(\lambda_1 - \lambda_2 + 2\vep)(\lambda_1+\lambda_2-1- 2\vep)u \, . \cr
}
}
The justification of \Frec\ follows by showing that \HHC\ is then 
satisfied for $\Delta_4^{(\vep)}$ having the eigenvalue \eigDf.

The coefficients $s,t,u$ in \Frec\ can be determined by analysing
the limit $x,\bx \to 0$,
\eqn\resstu{\eqalign{
r=1 \, , \quad &
s = - {\lambda_1-\lambda_2 + 2 \vep \over \lambda_1-\lambda_2 + \vep}\,
\sigma_{\lambda_1}(a,b) \, , \quad
t = - {\lambda_1-\lambda_2 \over \lambda_1-\lambda_2 + \vep}\,
\sigma_{\lambda_2 - \vep}(a,b) \, ,\cr
u = {}& {(\lambda_1+\lambda_2 -1)( \lambda_1+\lambda_2 -2 \vep)\over
(\lambda_1+\lambda_2 -1 - \vep)( \lambda_1+\lambda_2 -\vep)} \,
\sigma_{\lambda_1}(a,b)\, \sigma_{\lambda_2 - \vep}(a,b) \, .
}
}

By combining \Frec\ with \symF{a,b} we may then obtain
\eqnn\Fcom$$\eqalignno{
u^{-{1\over 2}} F_{\lambda_1\lambda_2}(a,b) = {}&
F_{\lambda_1-{1\over 2}\, \lambda_2-{1\over 2}}(a{+\half},b{+\half}) -
{\ell + 2 \vep\over \ell+\vep} \, \sigma_{\lambda_1}(a, b)\, 
F_{\lambda_1+{1\over 2}\, \lambda_2-{1\over 2}}(a{+\half},b{+\half}) \cr
&{}- {\ell \over \ell+\vep} \, \sigma_{\lambda_2-\vep}( a, b)\,    
\, F_{\lambda_1-{1\over 2}\, \lambda_2+{1\over 2}}(a{+\half},b{+\half}) \cr
&{}+ f_\Delta \, \sigma_{\lambda_1}(a, b)\,
\sigma_{\lambda_2-\vep}( a,  b)\,
F_{\lambda_1+{1\over 2}\, \lambda_2+{1\over 2}}(a{\pm\half},b{\mp\half}) \,,\cr
u^{-{1\over 2}} F_{\lambda_1\lambda_2}(a,b) = {}&
F_{\lambda_1-{1\over 2}\, \lambda_2-{1\over 2}}(a{\pm\half},b{\mp\half}) +
{\ell + 2 \vep\over \ell+\vep} \, \sigma_{\lambda_1}(\pm a, \mp b)\, 
F_{\lambda_1+{1\over 2}\, \lambda_2-{1\over 2}}(a{\pm\half},b{\mp\half}) \cr
&{}+ {\ell \over \ell+\vep} \, \sigma_{\lambda_2-\vep}(\pm a, \mp b)\, 
\, F_{\lambda_1-{1\over 2}\, \lambda_2+{1\over 2}}(a{\pm\half},b{\mp\half}) \cr
&{}+ f_\Delta \, \sigma_{\lambda_1}(\pm a, \mp b)\,
\sigma_{\lambda_2-\vep}(\pm a, \mp b)\,
F_{\lambda_1+{1\over 2}\, \lambda_2+{1\over 2}}(a{\pm\half},b{\mp\half}) \,,\cr
u^{-{1\over 2}}v F_{\lambda_1\lambda_2}(a,b) = {}&
F_{\lambda_1-{1\over 2}\, \lambda_2-{1\over 2}}(a{-\half},b{-\half}) -
{\ell + 2 \vep\over \ell+\vep} \, \sigma_{\lambda_1}(- a, - b)\, 
F_{\lambda_1+{1\over 2}\, \lambda_2-{1\over 2}}(a{-\half},b{-\half}) \cr
&{}- {\ell \over \ell+\vep} \, \sigma_{\lambda_2-\vep}(- a, - b)\,    
\, F_{\lambda_1-{1\over 2}\, \lambda_2+{1\over 2}}(a{-\half},b{-\half}) \cr
&{}+ f_\Delta \, \sigma_{\lambda_1}(- a, - b)\,
\sigma_{\lambda_2-\vep}(- a, - b)\,
F_{\lambda_1+{1\over 2}\, \lambda_2+{1\over 2}}(a{-\half},b{-\half}) \,, 
& \Fcom\cr
}$$
for
\eqn\fdel{
f_\Delta = {(\Delta-1)(\Delta + 2 -d)\over (\Delta-\half d)(\Delta-\half d
+1)} \, .
}
By iterating \Fcom\ we may rederive the result for
$u^{-1} F_{\lambda_1\lambda_2}(a,b)$ in terms of
$F_{\lambda_1{}'\lambda_2{}'}(a,b)$ for 
$(\lambda_1{}',\lambda_2{}') = (\lambda_1\pm 1,\lambda_2\pm 1), 
(\lambda_1\pm 1,\lambda_2),(\lambda_1,\lambda_2\pm 1),
(\lambda_1,\lambda_2)$ given in \DT. Furthermore, eliminating
$F_{\lambda_1-{1\over 2}\, \lambda_2-{1\over 2}}$ and
$F_{\lambda_1+{1\over 2}\, \lambda_2+{1\over 2}}$ from \Fcom\ leads to
\eqnn\Fll
$$\eqalignno{
(\ell + & \half d -1) u^{-{1\over 2}} \Big ( 
(\lambda_1+ b-\half)(\lambda_2 + b-\half - \vep) \big (
F_{\lambda_1\lambda_2}(a{-\half},b{+\half}) - v
F_{\lambda_1\lambda_2}(a{+\half},b{+\half})  \big ) \cr
\noalign{\vskip -2pt}
&\hskip 2cm{} + (\lambda_1- b-\half)(\lambda_2 - b-\half - \vep) \big (
F_{\lambda_1\lambda_2}(a{+\half},b{-\half}) -
F_{\lambda_1\lambda_2}(a{-\half},b{-\half})  \big ) \Big )\cr
= {}& (\ell+d-2){1\over \lambda_1}(\lambda_2- \half - \vep)
(\lambda_1-b-\half)(\lambda_1+b-\half) \,
F_{\lambda_1+{1\over 2}\, \lambda_2-{1\over 2}}(a,b) \cr
&{}+  \ell {1\over \lambda_2-\vep }(\lambda_1- \half)
(\lambda_2 -b-\half-\vep)(\lambda_2 +b-\half-\vep) \,
F_{\lambda_1-{1\over 2}\, \lambda_2+{1\over 2}}(a,b) \, , & \Fll
}$$
which is equivalent to \recurF.

Using \Frec\ and \FFrec\ with \rstu\  and \resstu\ gives
\eqnn\HHF$$\eqalignno{
\big (\H_2&  - \lambda_2 \, \H_1 + \lambda_2(\lambda_2-\vep)\, 
(x\bx)^{-{1\over 2}}
\big ) F_{\lambda_1\lambda_2}(a,b)  = u^{\lambda_2} \H_2 \, u^{-\lambda_2}
F_{\lambda_1\lambda_2}(a,b)  \cr
= {}& \sigma_{\lambda_2-\vep}(a,b)(2 \lambda_2 -1 - 2 \vep)\Big ( \ell \,
F_{\lambda_1-{1\over 2}\,\lambda_2+{1\over 2}}(a+\half, b+\half )  \cr
\noalign{\vskip -2pt} 
&\hskip 4cm  {}+   {(\Delta-1)(\Delta-2\vep)\over
\Delta-\vep} \, \sigma_{\lambda_1}(a,b) \, 
F_{\lambda_1+{1\over 2}\,\lambda_2+{1\over 2}}(a+\half, b+\half )  \Big ) \cr
= {}& 0 \quad \hbox{for} \quad \lambda_2 = \vep -b \quad \hbox{or}\quad
\lambda_2 = \vep-a \, . &\HHF 
 }
$$

There are further multi-term recurrence  relations  for conformal partial 
waves, for  fixed $\vep$ and $a,b$,
which extend the single variable results \idgp{a,b}. To obtain these we define  
\eqn\defF{\eqalign{
\F_0 = {}& {1\over x} + {1\over \bx} - 1 \, ,  \qquad
\F_1 = (1-x){\pr \over \pr x} + (1-\bx){\pr \over \pr \bx} -  a - b \, , \cr
\F_2 = {}& { x-\bx \over x\bx  } \, \big ( D_x - D_\bx \big ) \, , \cr
\F_3 = {}& \bigg (  (1-x){\pr \over \pr x} -  (1-\bx){\pr \over \pr \bx} 
-2\vep \,  {x+\bx -2 \over x- \bx} \bigg )  \big ( D_x - D_\bx \big ) \, , 
}
}
which satisfy the commutation relations with $ \Delta^{(\vep)}$,
generalising \Dff,
\eqn\comD{\eqalign{
\big [ \Delta^{(\vep)} , \F_0 \big ] ={}& - 2 \, \F_1 + 2(1+\vep ) \, \F_0 \, , 
\qquad
\big [ \Delta^{(\vep)} , \F_1 \big ] =  \F_2 - \F_0\,  \Delta^{(\vep)} - 
2ab  \, , \cr
\big [ \Delta^{(\vep)} , \F_2 \big ] = {}& 2\, \F_3   
+ 2(1+\vep)\, \F_2 \, , \cr
\big [ \Delta^{(\vep)} ,\F_3 \big ] = {}& - \F_0 \, \Delta_4{\!}^{(\vep)} 
- 4 \vep \, \F_3
+ \F_2 \big (\Delta^{(\vep)} - 2\vep(1+2\vep)  \big ) \, .
}
}
As a consequence of \comD\ it is sufficient to assume
\eqn\HFrec{
\F_i F_{\lambda_1\lambda_2} = 
r_i \, F_{\lambda_1\, \lambda_2- 1} +
s_i\, F_{\lambda_1-1\, \lambda_2}
+ t_i\, F_{\lambda_1+ 1\, \lambda_2}  +
u_i\, F_{\lambda_1\, \lambda_2+ 1} + w_i \, F_{\lambda_1\lambda_2} \, ,
}
Imposing \comD\ then leads to
\eqnn\rstuw
$$\eqalignno{
r_1 = {}& \lambda_2 \, r_0 \, , \quad s_1=(\lambda_1 +\vep) s_0 \, , \quad
t_1 = - (\lambda_1- 1 - \vep) t_0 \, , \quad
u_1 = - (\lambda_2- 1 - 2\vep) u_0 \, , \cr
r_2 = {}& (\lambda_1-\lambda_2)(\lambda_1+\lambda_2-1)r_0 \, , \quad
s_2 = - (\lambda_1-\lambda_2+ \vep )(\lambda_1+\lambda_2-1)s_0 \, ,\cr
t_2 ={}& - (\lambda_1-\lambda_2)(\lambda_1+\lambda_2-1 -2\vep )t_0  \, , \quad
u_2 =  (\lambda_1-\lambda_2+2\vep )(\lambda_1+\lambda_2-1 -2\vep )u_0 \, , \cr
& r_3 = - \lambda_2 (\lambda_1-\lambda_2)(\lambda_1+\lambda_2-1)r_0 \, , \cr
& s_3 = (\lambda_1 +\vep)  (\lambda_1-\lambda_2+ \vep )
(\lambda_1+\lambda_2-1)s_0 \, , \cr
& t_3 = - (\lambda_1- 1 - \vep) (\lambda_1-\lambda_2)
(\lambda_1+\lambda_2-1 -2\vep )t_0  \, , \cr
& u_3 = (\lambda_2- 1 - 2\vep) (\lambda_1-\lambda_2+2\vep )
(\lambda_1+\lambda_2-1 -2\vep )u_0 \, , &\rstuw
}
$$
and also $w_i$ is determined giving
\eqn\resw{\eqalign{
w_1 = {}& (1+\vep)w_0\, , \qquad w_3=-(1+\vep)w_2 \, , \cr
w_0 = {}& - (c_{\lambda_1\lambda_2}+ 2 \vep) \, {ab \over 2d_{\lambda_1\lambda_2}}\, ,
\quad w_2 = - c_{4,\lambda_1\lambda_2} \, {ab \over 2d_{\lambda_1\lambda_2}} \, ,
}
}
where
\eqn\resd{
d_{\lambda_1\lambda_2} = - \quar \big (  c_{4,\lambda_1\lambda_2}
- c_{\lambda_1\lambda_2}{\!}^2 - 2\vep \, c_{\lambda_1\lambda_2}\big )
= \lambda_1(\lambda_1-1)(\lambda_2- \vep)
(\lambda_2 - 1 - \vep) \, .
}
The coefficients are the determined by taking in \rstuw
\eqn\resstuiw{\eqalign{
r_0 = {}&  {\lambda_1-\lambda_2 + 2 \vep \over \lambda_1-\lambda_2 + \vep}\,
 \, , \qquad
s_0 =  {\lambda_1-\lambda_2 \over \lambda_1-\lambda_2 + \vep}\, \, ,\cr
t_0 = {}& {(\lambda_1+\lambda_2 -1)( \lambda_1+\lambda_2 -2 \vep)\over
(\lambda_1+\lambda_2 -1 - \vep) ( \lambda_1+\lambda_2 -\vep)} \, 
{\lambda_1-\lambda_2 + 2 \vep \over \lambda_1-\lambda_2 + \vep}\,\beta_{\lambda_1}\, ,
\cr
u_0 = {}& {(\lambda_1+\lambda_2 -1)( \lambda_1+\lambda_2 -2 \vep)\over
(\lambda_1+\lambda_2 -1 - \vep) ( \lambda_1+\lambda_2 -\vep)} \, 
{\lambda_1-\lambda_2 \over \lambda_1-\lambda_2 + \vep}\, \beta_{\lambda_2-\vep} \, .
}
}

Explicitly the result for $\F_0 \,  F_{\lambda_1\lambda_2}$ becomes
\eqn\recurF{\eqalign{
\F_0  \, F_{\lambda_1\lambda_2} = {}&
{{\ell+2\vep}\over {\ell+ \vep}} \, F_{\lambda_1\lambda_2{-1}} +
 {\ell \over \ell +  \vep }\, F_{\lambda_1{-1}\lambda_2} 
 - (c_{\lambda_1\lambda_2}+ 2 \vep) \, {ab \over 2d_{\lambda_1\lambda_2}} \, 
 F_{\lambda_1\lambda_2} \cr
&{}+ f_\Delta \bigg (
{\ell + 2\vep \over \ell + \vep } \, \beta _{\lambda_1}  
F_{\lambda_1{+1}\lambda_2}(x,\bx) +  {\ell \over \ell +  \vep } \, 
\beta_{\lambda_2 - \vep} F_{\lambda_1\lambda_2{+1}}\bigg )\, ,
\cr}
}
with $f_\Delta$ given by \fdel. The result for $\F_2 \,  F_{\lambda_1\lambda_2}$
vanishes for $\ell=0$ so that
\eqn\lzer{
\big ( D_x - D_\bx \big ) F_{\lambda\,\lambda}(x,\bx) = 0 \, .
}

There are also second order differential operators relating conformal 
partial waves for dimensions differing by 2. Defining, in terms of $\F_2$
in  \defF,
\eqn\Epdef{
\E_+ = \Big ({x\bx \over x-\bx}\Big )^2 \F_2 
= {x\bx\over x-\bx} \, \big ( D_x - D_\bx \big )  \, ,
}
then we may verify that
\eqn\EpD{
\E_+ \, \Delta^{(\vep)} = 
\big (  \Delta^{(\vep+1)} + 2 + 2\vep \big )\, \E_+ \, .
}
For $\bx\to 0$ followed by $x \to 0$, 
$\E_+ \sim \bx \big  (x^2 {\pr^2\over \pr x^2} - \bx^2 
{\pr^2\over \pr\bx^2}\big)$
so that, assuming this limit,
$\E_+ x^{\bl_1}\bx^{\bl_2} \sim (\lambda_1-\lambda_2)(\lambda_1+\lambda_2-1)
\, x^{\bl_1}\bx^{\bl_2+1}$. Hence, with the normalisation 
of $F^{(\vep)}{\!}_{\lambda_1\lambda_2}$, determined by \cell,
\eqn\EpF{
\E_+ F^{(\vep)}{\!}_{\lambda_1\lambda_2}(x,\bx) =
{\lambda_1 + \lambda_2-1\over 2(2\vep+1)} \,
(\lambda_1-\lambda_2)(\lambda_1-\lambda_2+ 2\vep ) \,
F^{(\vep+1)}{\!}_{\lambda_1\,\lambda_2+1}(x,\bx) \, .
}
As a special case\foot{To solve this equation it is more
convenient to revert to the $u,v$ variables when
$$
\E_+ =  u \bigg (  v(1-v) {\pr^2 \over \pr v^2} -  2uv {\pr^2 \over \pr u \pr v}
- u^2 {\pr^2 \over \pr u^2} + (a+b+1) \Big ( (1-v) {\pr \over \pr v} 
- u {\pr \over \pr u} \Big ) - ab \bigg ) \, .
$$
Then $\E_+ G(u,v) =0$ has solutions, analytic for $1-v \approx 0$, 
$u^\lambda F(\lambda+a, \lambda+b;2\lambda;1-v)$ for any $\lambda$.
This is in accord with the conformal partial wave when $\ell=0$ 
given by \Gzero.} $\E_+ F^{(\vep)}{\!}_{\lambda \lambda} = 0$ which is
equivalent  to \lzer.
 
By combining \EpD\ with \pid\ we may obtain a related operator which
has the reverse effect
\eqn\Emdef{
\E^{(\vep)}_{\,\,-} = \bigg ( {x\bx \over x-\bx} \bigg )^{\! 2 \vep - 2}
\big ( D_x - D_\bx \big ) \, 
\bigg ( {x\bx \over x-\bx} \bigg )^{\! - 2\vep + 1} \, ,
}
such that
\eqn\EmD{
\E^{(\vep)}_{\,\,-} \Delta^{(\vep)} 
= \big (  \Delta^{(\vep-1)} - 2\vep  \big ) \, \E^{(\vep)}_{\,\,-} \, .
} 
Similarly to \EpF\ we then have
\eqn\EmF{
\E^{(\vep)}_{\,\,-}  F^{(\vep)}{\!}_{\lambda_1\lambda_2}(x,\bx) =
2(2\vep-1) \,  (\lambda_1+\lambda_2- 2\vep ) \,
F^{(\vep-1)}{\!}_{\lambda_1\,\lambda_2-1}(x,\bx) \, .
}
It is crucial that both $\E_+$ and $\E^{(\vep)}_{\,\,-}$, defined in \Epdef\
and \Emdef, are symmetric under $x\leftrightarrow \bx$. It is easy to see from
\Casz
\eqn\EECas{
 \E^{(\vep+1)}_{\,\,-} \E_+ = \Delta_4{\!}^{(\vep)} \, .
}

From the result from $\F_2 F_{\lambda_1 \lambda_2}$,
\eqn\recurDDF{\eqalign{
\hskip - 1cm & \bigg ( {x-\bx \over x \bx}  \bigg )^{\! 2} \;
{\ell+1+ \vep \over  2(2\vep+1)} \;
F^{(\vep+1)}{\!}_{\lambda_1\lambda_2}(x,\bx)\cr
& \!\! ={} 
F^{(\vep)}{\!}_{\lambda_1\,\lambda_2{-2}}(x,\bx) - 
F^{(\vep)}{\!}_{\lambda_1{-1}\,\lambda_2-1}(x,\bx) \cr
&   {} - ( \lambda_1 + \lambda_2 - 2 - 2 \vep )(\ell+1+ \vep) \,
{ab \over 2 {d}_{\lambda_1\lambda_2-1}} \, 
F^{(\vep)}{\!}_{\lambda_1\lambda_2-1}(x,\bx) \cr
&    {}
-{(\lambda_1+\lambda_2-1- 2\vep )(\lambda_1+\lambda_2 -2 -2\vep)\over
(\lambda_1+\lambda_2-1-  \vep )(\lambda_1+\lambda_2 -2 -  \vep)}\,
\big ( \beta _{\lambda_1}
F^{(\vep)}{\!}_{\lambda_1{+1}\,\lambda_2-1}(x,\bx) - 
\beta_{\lambda_2 -1-  \vep} 
F^{(\vep)}{\!}_{\lambda_1\,\lambda_2}(x,\bx)\big )  . \cr}
}
An equivalent result was obtained in \DT, which, along with \recurF,
was derived using manipulations based on a series expansion
in terms of Jack polynomials.\foot{The analogous formula in the compact case
was given in \Vretare\ and is rederived in appendix B.} 

Further recurrence relations for conformal partial waves 
can be derived by using the second order operator
\eqn\defDD{ 
\D^{(\vep)} = (x\bx)^{-{1\over 2}} \, \H_2 
= {1\over (x-\bx)^\vep} \, {\pr^2\over \pr x \pr\bx} \, (x-\bx)^\vep \, , 
}
with $\H_2$ given in \HHH, or equivalently
\eqn\Drt{
4\, \D^{(\vep)} = {\pr^2\over \pr t^2} - {\pr^2 \over \pr r^2} -
{2\vep \over r} \, {\pr \over \pr r}  \quad \hbox{for} \quad
t=\half(x+\bx),\, r = \half (x-\bx) \, .
}
It is then clear that  $\D^{(\vep)}{\hat F}=0$ is just the spherically 
symmetric, or $S$-wave, projection of the wave equation in $2(\vep+1)$ 
dimensions, although here we require only solutions even in $r$. 
From its definition \defDD\ $\D^{(\vep)}$ is invariant under reflection 
$x\to -x, \bx\to - \bx$. $\D^{(\vep)}$ also  satisfies, analogous to \pid, 
\eqn\piD{ 
\D^{(\vep)} {1\over (x-\bx)^{2\vep-1}} = {1\over (x-\bx)^{2\vep-1}} \, 
\D^{(-\vep+1)} \, . 
} 

The wave equation is invariant under conformal 
transformations which preserve the spherical symmetry. In this case for
\eqn\xxpr{
x' = {\alpha x + \beta \over \gamma x + \de } \, , \quad
\bx' = {\alpha \bx + \beta \over \gamma \bx + \de } \, , \qquad
\alpha \de - \beta \gamma = 1 \, ,
}
so that $\pmatrix{\alpha & \beta \cr \gamma & \delta } \in Sl(2, \Bbb R)$, 
then
\eqn\Dpr{
\D^{(\vep)} \toinf{x\to x', \bx\to \bx'}
\D'{}^{(\vep)} = \Omega^{2+ \vep} \; \D^{(\vep)} \, \Omega^{ -\vep} \, ,
}
where
\eqn\Ome{
\Omega = ( \gamma x + \de) (\gamma \bx + \de ) = {1\over (\alpha - \gamma x')
(\alpha - \gamma \bx')}  \, .
}

With $u,v$ defined in \defxz, we may obtain
\eqn\DDela{
\Delta^{(\vep)}(a,b+1) \, u^{-b+1+ \vep}  \D^{(\vep)} u^{b- \vep }   
=  u^{-b+1+ \vep}  \D^{(\vep)} u^{b- \vep }  \Delta^{(\vep)} (a,b) \, ,
}
Hence we may obtain
\eqn\relFa{
u\, \D^{(\vep)}\,  u^{b- \vep } F_{\lambda_1\lambda_2}(a,b;x,\bx)
=  (\lambda_1+b) (\lambda_2+b - \vep) \, u^{b- \vep }
F_{\lambda_1\lambda_2}(a,b+1,x,\bx) \, .
}
Conversely using \gveq
\eqn\relFb{
uv \, \D^{(\vep)} \, u^{- b- \vep }v^{a+b}
F_{\lambda_1\lambda_2}(a,b;x,\bx) 
= (\lambda_1-b) (\lambda_2-b -\vep)  \, u^{- b- \vep }v^{a+b} 
F_{\lambda_1\lambda_2}(a,b-1;x,\bx) \, .
}
Of course there are similar results for $a\leftrightarrow b$.
Consistency \relFa\ and \relFb\ requires the fourth order 
eigenvalue equation,
\eqn\CasF{
u^{2+\vep} \D^{(\vep)} v \D^{(\vep)}  u^{-\vep}\,
F_{\lambda_1\lambda_2}(0,0;x,\bx) =  d_{\lambda_1\lambda_2} \,
F_{\lambda_1\lambda_2}(0,0;x,\bx)  \, .
}
for $ d_{\lambda_1\lambda_2} $ as in \resd.

The right hand side of \relFa\ vanishes when $\lambda_2 = \vep - b$.
Using 
$\D^{(\vep)} u^{N+1} \D^{(\vep)}{}^N = u^N\, \D^{(\vep)}{}^{N+1} u$
and applying \relFa\ repeatedly this may be extended to
\eqn\DFh{ 
\D^{(\vep)}{}^{N}\big( (x\bx)^{b+N-1- \vep}
F^{(\vep)}{\!}_{\lambda_1\lambda_2} (x,\bx) \big ) \big |_{
\lambda_2 = \vep -b +1 -N} = 0 \, , \quad  N=1,2 \dots \, . }
This result leads to additional constraints on  conformal partial waves 
for  particular values of the twist $\Delta-\ell$. For $N=1$ \DFh\ coincides
with the vanishing condition in \HHF.

Corresponding to \relFa\ and \relFb\ there are analogous single variable
relations
\eqn\gpb{\eqalign{
x {\d \over \d x} \, x^b g_p(a,b;x) = {}& (p+b) \, x^b g_p(a,b+1;x) \,,  \cr
x(1-x) {\d \over \d x} \, x^{-b}(1-x)^{a+b} g_p(a,b;x) = {}&
(p-b) \, x^{-b}(1-x)^{a+b} g_p(a,b-1;x) \, .
}
}

\newsec{Even Dimensional Results}

For even $d$ the solutions are all expressible in terms of the
symmetric, antisymmetric combinations of  $g_p,g_q$, 
as defined in \hyper, for suitable $p,q$,
\eqn\Fgg{\eqalign{
\F^{\pm}{\!\!}_{pq}(x,\bx) = {}& g_{p}(x) \, g_{q}(\bx) \pm 
g_{p}(\bx) \, g_{q}(x) \limu{\bx\to 0,x\to 0} (x\bx)^{q} \, x^{p-q} \, , \quad 
\hbox{for}  \quad p-q = 1,2, \dots \, ,
}
}
Assuming $g_p(\bx)$ transforms in the conjugate fashion to $g_p(x)$
in \gab\ then from \Fgg
\eqn\Fxp{\eqalign{
\F^{\pm}{\!\!}_{pq}(x,\bx) = {}& (-1)^{p-q}
v^{-b} \F^{\pm}{\!\!}_{pq}(x',\bx') \, \big |_{a\to -a} \cr
= {}& (-1)^{p-q}
v^{-a} \F^{\pm}{\!\!}_{pq}(x',\bx') \, \big |_{b\to -b} \, ,
}
}
which is sufficient to verify \symF{a,b} in each case. The identity \symFF\
follows in each case from $\F^{\pm}{\!\!}_{pq} = \pm \F^\pm{\!\!}_{qp}$.

For $d=2$, or $\vep=0$, $\Delta^{(0)}$ becomes a sum of single
variable operators so that it is trivial to separate variables and the
conformal partial waves are just, since from \cell\ $c_\ell{\!}^{(0)} = \half$
for $\ell \ne 0$,
\eqn\Fzero{
2 \,  F^{(0)}{\!}_{\lambda_1\lambda_2}(x,\bx) = 
\F^+{\!\!}_{\lambda_1\, \lambda_2}(x,\bx)\, ,
}
agreeing of course with \Gtwo.

For  $\vep=1$
\eqn\two{
(\lambda_1 - \lambda_2 +1) \, F^{(1)}{\!}_{\lambda_1\lambda_2}(x,\bx) =
{x\bx \over x - \bx} \;  \F^-{\!\!}_{\lambda_1\, \lambda_2-1}(x,\bx)\, .
}
With the operator $\E_+$ defined by \Epdef, and using the eigenvalue equation
\defg, it is very easy to verify that \two\ and \Fzero\ are in accord with
\EpF\ raising $\vep$ by one.

To obtain results for $\vep=2$  using \EpF\ directly, which becomes
\eqn\EpTwo{
\E_+ F^{(1)}{\!}_{\lambda_1\lambda_2}(x,\bx) = {\ts{1\over 6}}
(\lambda_1 + \lambda_2-1) \,
(\lambda_1-\lambda_2)(\lambda_1-\lambda_2+ 2 ) \,
F^{(2)}{\!}_{\lambda_1\,\lambda_2+1}(x,\bx) \, ,
}
it is necessary to calculate the action of $\E_+$, as given
by \Epdef, on $F^{(2)}{\!}_{\lambda_1\lambda_2}$. Using \defg\ again we get
\eqn\Efour{\eqalign{
& (\lambda_1 - \lambda_2 +1)\; \E_+ F^{(1)}{\!}_{\lambda_1\lambda_2}(x,\bx) \cr
&{} = -  \bigg ({x\bx \over x - \bx}  \bigg)^{\! 3} \, \bigg \{
2 \Big ( (1-x){\pr \over \pr x} + (1-\bx){\pr \over \pr \bx} - a -b 
\Big )\,  \F^-{\!\!}_{\lambda_1\, \lambda_2-1}(x,\bx) \cr
\noalign{\vskip -4pt}
&\hskip 2.7cm{} + (\lambda_1 - \lambda_2 +1)(\lambda_1 + \lambda_2  -2)
\Big ( {1\over x} - {1\over \bx} \Big )   \, 
\F^+{\!\!}_{\lambda_1\, \lambda_2-1}(x,\bx) \bigg \}  \, .
}
}
As a consequence of \idgp{a,b} we have
\eqna\rgg$$\eqalignno{
& \Big ( {1\over x} - {1\over \bx} \Big )   \,
\F^+{\!\!}_{pq}(x,\bx) = \F^-{\!\!}_{p-1\, q}(x,\bx) 
- \F^-{\!\!}_{p\, q-1}(x,\bx) +  (\alpha_p - \alpha_q) \, \F^-{\!\!}_{pq}(x,\bx)\cr
&\hskip 4cm {} + \beta_p \,  \F^-{\!\!}_{p+1\, q}(x,\bx) -
\beta_q \,  \F^-{\!\!}_{p\, q+1}(x,\bx) \, , & \rgg{a}\cr
&\Big ( (1-x){\pr \over \pr x} + (1-\bx){\pr \over \pr \bx} - a -b
\Big )\,  \F^-{\!\!}_{pq}(x,\bx) \cr
&\hskip 2cm {} =  p\, \F^-{\!\!}_{p-1\, q}(x,\bx)   
+ q\, \F^-{\!\!}_{p\, q-1}(x,\bx) + (\alpha_p + \alpha_q  ) 
\, \F^-{\!\!}_{pq}(x,\bx)\cr
&\hskip 4cm {} -(p-1) \beta_p \,  \F^-{\!\!}_{p+1\, q}(x,\bx) -
(q-1) \beta_q \,  \F^-{\!\!}_{p\, q+1}(x,\bx) \, ,  & \rgg{b}
}
$$
for $\alpha_p, \beta_p$ is given by \idab.

Using \rgg{a,b} in \Efour\ then gives from \EpTwo\ 
\eqn\four{\eqalign{
{\ts {1\over 6}} (\lambda_1 &{} - \lambda_2 +3)(\lambda_1 - \lambda_2 +2)
(\lambda_1 - \lambda_2 +1)\, F^{(2)}{\!}_{\lambda_1\lambda_2}(x,\bx) \cr
=  {}& \bigg ({x\bx \over x - \bx}  \bigg)^{\! 3} \, \bigg \{
(\lambda_1-\lambda_2+1)\, \F^-{\!\!}_{\lambda_1 \, \lambda_2-3}(x,\bx) \cr
\noalign{\vskip-6pt}
&\hskip 3cm{} 
- (\lambda_1-\lambda_2+3)\, \F^-{\!\!}_{\lambda_1-1 \,\lambda_2-2}(x,\bx)\cr
&- {\lambda_1+\lambda_2-4\over \lambda_1+\lambda_2-2} \,
\Big ((\lambda_1-\lambda_2+1) \beta_{\lambda_1} \,
\F^-{\!\!}_{\lambda_1+1 \,\lambda_2-2}(x,\bx) \cr
\noalign{\vskip-6pt}
&\hskip 4cm{} - (\lambda_1-\lambda_2+3) \beta_{\lambda_2-2}\,
\F^-{\!\!}_{\lambda_1\,\lambda_2-1}(x,\bx) \Big ) \cr
&{}- (\lambda_1+\lambda_2-4)(\lambda_1-\lambda_2+1)(\lambda_1-\lambda_2+3)
\, {ab \over 2 d_{\lambda_1 \lambda_2}}\,
\F^-{\!\!}_{\lambda_1\,\lambda_2-2}(x,\bx) \bigg \} \, ,
}
}
with $ d_{\lambda_1 \lambda_2} = 
\lambda_1(\lambda_1-1)(\lambda_2-2)(\lambda_2-3)$.
The right hand side of \four\ vanishes when $\lambda_1-\lambda_2=-1,-2,-3$.
That the expression
for $F^{(2)}{\!}_{\lambda_1\lambda_2}(x,\bx)$ given by \four\ is finite
when $x=\bx$ is verified in appendix C.

For particular twist the solutions simplify as expected from \DFh.
For the case when $N=1$ we require solutions of
\eqn\lowt{
\D^{(\vep)} {\hat F} ( x,\bx ) = 0 \, ,
}
for ${\hat F} ( x,\bx )$ symmetric in $x,\bx$. 
The general solutions of \lowt\
involve an unconstrained single variable function $f$. For even $\vep$ the 
solutions, which may be found using \piD, are quite simple. In the first 
three cases 
\eqn\evenF{ 
  {\hat F}(x,\bx) = \cases{f(x)+f(\bx) \, , & $\vep= 0 \, , 
$\cr\noalign{\vskip 2pt} {f(x)-f(\bx)\over x - \bx} \, , & $\vep =1 \, , 
$\cr \noalign{\vskip 2pt} { f(x)- f(\bx) - {1\over 2}(x-\bx) 
(f'(x)+f'(\bx) )\over (x - \bx)^3} \, , & $ \vep =2 \, . $ } } 
It is easy 
to verify that ${\hat F}(x,\bx)$ is regular as $x\to \bx$ so long as $f$ is 
a smoothly differentiable function, so that from \evenF
\eqn\Fequal{
{\hat F}(x,x) =  2f(x), \,  f'(x), \, 
- {1\over 12}f'''(x) \qquad \hbox{for}   \qquad \vep=0,1,2 \, .
}

It is straightforward to verify how the solutions obtained above for
$\vep=0,1,2$ connect with the special cases
given by \evenF. For $\vep=0$ we have from \Fzero
\eqn\fzero{
F^{(0)}{\!}_{\lambda_1\lambda_2}(x,\bx) \big |_{\lambda_2 = -b} =
(x\bx)^{-b} \big ( f(x) + f(\bx) \big ) \, , \quad
f(x) = x^b g_{\lambda_1}(x) \, ,
}
and for $\vep=1$ from \two
\eqn\ftwo{
(\lambda_1- \lambda_2+1) F^{(1)}{\!}_{\lambda_1\lambda_2}(x,\bx) 
\big |_{\lambda_2 = 1-b} = {(x\bx)^{1-b}\over x-\bx}\, 
\big ( f(x) - f(\bx) \big ) \, , \quad
f(x) = x^{b} g_{\lambda_1}(x) \, .
}
For $\vep=2$ we require
\eqn\ffour{\eqalign{
{\ts {1\over 6}} (\lambda_1 &{} - \lambda_2 +3)(\lambda_1 - \lambda_2 +2)
(\lambda_1 - \lambda_2 +1)\, 
F^{(2)}{\!}_{\lambda_1\lambda_2}(x,\bx)\big |_{\lambda_2 =2-b} \cr
= {}& {(x\bx)^{2-b} \over (x - \bx)^3}\,  \Big ( 
f(x) - f(\bx) - \half( x-\bx) \big ( f'(x) + f'(\bx) \big ) \Big ) \, .
}
}
From \four, using the expansion of $F(a-b-1,-1;-2b-2;x)$, this requires
\eqn\ffp{
f(x) - \half x f'(x) = (\lambda_1- 1 + b) \, x^{b+1} g_{\lambda_1}(x) \, ,
}
and 
\eqnn\fpeq
$$\eqalignno{
x^{-b-1} \half f'(x) = {}& - (\lambda_1 + 1 +b ) \, g_{\lambda_1-1}(x)
+ \half (\lambda_1 - 1 +b )\bigg ( -1 + 
{a(b+2) \over \lambda_1(\lambda_1-1)}\bigg )\, g_{\lambda_1}(x) \cr
&{}-  (\lambda_1 - 1 +b ) \, {\lambda_1-2 -b \over \lambda_1 - b} \,
g_{\lambda_1+ 1}(x) \, .  & \fpeq 
}
$$
\ffp\ is satisfied by taking
\eqn\fsol{
f(x) = - 2x^{b+1} \, g_{\lambda_1}(a,b-1;x) \, .
}
This also satisfies \fpeq, as may be verified by using \ffp\ to eliminate 
$f'(x)$ and then using \idgp{a} for $g_{\lambda_1}(x)/x$ and also
\eqn\hidgp{
{1\over x}\, g_p(a,b-1;x) = g_{p-1}(a,b;x) +
{\hat \alpha}_p(a,b) \, g_p(a,b;x) 
+ {\hat \beta}_p(a,b) \, g_{p+1}(a,b;x) \, , }
where
\eqn\hidab{\eqalign{
{\hat \alpha}_p(a,b) = {}& - {a(p+b-1) \over 2p(p-1)} \, , \cr
{\hat \beta}_p(a,b) = {}& - {(p+a)(p+b)(p-a)(p+b-1) \over 4p^2(2p+1)(2p-1)} \, .
}
}
\hidgp\ may be derived from \threeg\ so that  ${\hat \alpha}_p(a,b) = 
- \sigma_p(a,b-1)+ \sigma_{p-{1\over 2}}(-a-\half,b-\half)$, ${\hat \beta}_p(a,b) = 
- \sigma_p(a,b-1)\, \sigma_{p+{1\over 2}}(-a-\half,b-\half)$.

\newsec{Results for One and Three Dimensions}

For odd $d$ no such explicit formulae as for even $d$ described above are 
known. When $d=1$ or $\vep =- \half$, corresponding to conformal quantum 
mechanics, conformal partial waves are rather trivial. The conformal group
$SO(1,2)$ acting on four points leaves just one invariant instead
of the two given by $u,v$. The necessary constraint is obtained by
imposing $x = \bx$. In this case  from \defD
\eqn\oned{
\Delta^{(-{1\over 2})}(a,b) F(t) = \half D_t(2a,2b) F(t) +{\rm O}\big (
(x-\bx)^2 \big ) \, , \qquad t= \half(x+\bx) \, ,
}
and
\eqn\onxx{
\Delta^{(-{1\over 2})}(a,b) ( x-\bx)^2 = 
2 \big ( 2 - (a+2)(b+2)t \big ) \,  ( x-\bx)^2 \, .
}
Hence \eigF\ may be restricted to $x=\bx$ for $\vep=-\half $ 
and the conformal partial waves are given in \GF\ for this case
with $\ell=0$ by
\eqn\Fone{
F^{(-{1\over 2})}{\!}_{\lambda\lambda}(x,x) = g_{2\lambda}(2a,2b;x) 
= x^\Delta F(\Delta+2a,\Delta+2b;2\Delta;x) \, .
}
The requirement that $\ell = \lambda_1- \lambda_2 =0$ follows since
for the conformal group $SO(1,2)$ the representations are determined 
just by the scale dimension  $\Delta$. 

For $d=3$ or $\vep=\half$ solutions of  \lowt\ of a similar form to \evenF\ 
are not possible,
reflecting the difference between the properties  of the wave equation in
even and odd dimensions.\foot{For $d=2,4$ the solutions of the
spherically symmetric wave equation are $f(t+r) + g(t-r)$, 
${1\over r}(f(t+r) + g(t-r))$ where the dependence on $t\pm r$ reflects
Huygens principle. The solutions in \evenF\ are obtained by requiring
them to be even in $r$, so that $g=\pm f$. Huygens principle does not hold
in odd dimensions.}
However when $d=3$  solutions of the wave equation can be found in terms of
the Bateman transform \Bate, which is related to twistor transforms. Hence 
we may obtain a solution of $\D^{({1\over 2})}{\hat F}=0$ in terms of the 
equivalent integral expressions
\eqn\thF{ 
{\hat F}(x,\bx) = {1\over \pi} \int_0^\pi \! \d s \; f(X) 
= {1\over 2\pi i} \oint_{|z|=1}\!\! {\d z \over z} \; f(X)  \, , } 
for 
\eqn\defX{ 
X= \cos^2 \! s \; x + \sin^2\! s \; \bx = t + r \, \half (z+z^{-1} ) \, . } 
Clearly from \thF\
\eqn\hFf{
{\hat F}(x,x) = f(x) \, .
}
To verify \thF\ provides a solution of \DFh, with $N=0$, it is sufficient 
to use 
\eqn\DFeq{ 
\D^{({1\over 2})} f(X) = -{1\over 4}\, {\pr \over \pr s} 
\bigg ( {\sin 2s \over x-\bx} \, f'(X) \bigg) \sim 0 \, , } 
where $\sim$ denotes equality up to terms which vanish under integration 
in \thF. For this to be valid $f(X)$ should have no branch point at $X=0$.
Subject to this condition we suppose that \thF\ is in fact a general 
solution but this depends on global issues not considered here. 

The integral representation \thF\ satisfies various identities.
As a consequence of \Dpr\ for the transformation \xxpr,
assuming \hFf\ and ${\hat F}(x,\bx)$ given by \thF, we must have
\eqn\thFp{ 
\Omega^{-{1\over 2}}
{\hat F}(x',\bx') = {1\over \pi} \int_0^\pi \! \d s \; 
{1\over \gamma X + \de} \,
f\Big ( {\alpha X + \beta \over \gamma X + \de} \Big ) \, .
}
To verify this we may note that, with $x',\bx'$ given by \xxpr\ and $\Omega$ by \Ome,
\eqn\mobX{
 {\alpha X + \beta \over \gamma X + \de} = 
\cos^2 \! s' \; x' + \sin^2\! s' \; \bx' \, , \qquad
{1\over \gamma X + \de} = {1\over \Omega^{1\over 2}} \, {\d s' \over \d s} \, ,
}
for
\eqn\ssp{
\cos^2 \! s' = { k \, \cos^2 \! s \over k  \cos^2 \! s + k^{-1} \sin^2  \! s} \, ,
\quad {\d s' \over \d s}  = 
{ 1 \over k  \cos^2 \! s + k^{-1} \sin^2  \! s} \, , \quad
k = \bigg ( {\alpha - \gamma \, \bx' \over \alpha - \gamma \, x' }
\bigg )^{\!{1\over 2}} \, .
}
A corollary of \thFp\ for the transformation $x,\bx \to 1/x,1/\bx$ gives
\eqn\invx{
{1\over \pi} \int_0^\pi \! \d s \; f(X)  =  
u^{1\over 2} \, {1\over \pi} \int_0^\pi \! \d s \; 
{1\over X} \, f \Big ({u \over X} \Big )  \, .
} 
Also a special case ensures that, assuming \thF,
\eqn\invv{
{\hat F}(x',\bx') = v^{1\over 2} \, 
{1\over \pi} \int_0^\pi \! \d s \;
{1\over 1- X} \, f \Big ({X \over X -1 } \Big )  \, , \qquad
x'= {x \over x-1} \, , \ \bx'= {\bx \over \bx-1} 
}

For functions $f(X)$ expressible as a Laurent series then  ${\hat F}(x,\bx)$ 
given by \thF\ is a corresponding sum of Legendre polynomials using
\eqn\Leg{
{1\over \pi} \int_0^\pi \! X^n \, \d s = u^{{1\over 2}n} P_n(\sigma) \, , 
}
where $\sigma$ is defined in \defsi\  so that \defX\ becomes
$X= (x\bx)^{1\over 2}( \sigma + \sqrt{\sigma^2-1} \, \cos 2s )$. 
The relation \invx\ is automatic if we identify
\eqn\idL{
P_n(\sigma) = P_{-n-1}(\sigma) \, .
}

To construct conformal partial waves we use, for $X$ defined by \defX,
\eqnn\Dfeq
$$\eqalignno{
\Delta^{({1\over 2})}& ( a , b) f(X)\cr
= {}& D_X( a + b - \half , 1 ) \, f(X)\cr
&{}+ {1\over 4} {\pr \over \pr s} 
\bigg ( \sin 2s \Big ( {x\bx \over x-\bx} \, (2-x-\bx)\, f'(X)
+ ( x-\bx )\big ( X f'(X)  + ( a+ b - \half ) f(X) \big ) \Big ) \bigg ) 
\cr \noalign{\vskip 2pt} &{} - ( a-\half ) ( b - \half ) \, (x+ \bx ) 
f(X) \, , & \Dfeq } $$ 
and hence
\eqn\Done{\eqalign{
\Delta^{({1\over 2})}& ( a , b) \big ( (u/X)^q f(X) \big ) \cr
&{} \sim (u /X)^q \big ( 
D_X (q + a+b -\half,1-q) f(X) + q(q- 2)\, f(X) \cr
& \hskip 3cm {}- (q+a-\half)(q+b-\half)  \, (x+ \bx ) f(X) \big ) \, .
} 
}
Writing then
\eqn\fint{
F^{({1\over 2})}{\!}_{\lambda_1\lambda_2}(x,\bx)  
= {1\over \pi} \int_0^\pi \! \d s \; \Big ( { u \over X}
\Big )^{\lambda_2} \, f(X) \quad \hbox{for} \quad \lambda_2 
= \half - b \, , 
}
the eigenvalue equation \eigF\ becomes in this case using \Done
\eqn\eigD{ 
D_X ( a , b+ \half ) f(X) = \lambda_1 (\lambda_1 -1 ) \, f(X) \, . } 
This is exactly of the form in \defg\ so that \eigD\ may be solved by 
taking 
\eqn\Solfs{ 
f(X) = g_{\lambda_1}(a,b+\half;X) \, , 
}
with $g_{\lambda_1}$ given explicitly by \hyper.
With this result we then have
\eqn\spF{\eqalign{
F^{({1\over 2})}{\!}_{\lambda_1\lambda_2}(a,b;x,\bx) \, \big |_{\lambda_2
= {1\over 2} - b} ={}&  {1\over \pi} \int_0^\pi \! \d s \; \Big ( { u \over X}
\Big )^{{1\over 2} - b} \,  g_{\lambda_1}(a,b+\half;X) \cr
= {}& u^{{1\over 2} - b} \, {1\over \pi} \int_0^\pi \! \d s \; 
 X^{\ell} F(\bl_1+a , \ell+1 ; 2\bl_1 ; X) \, .
}
}
Since $F(\bl_1+a , \ell+1 ; 2\bl_1 ; X) = (1-X)^{-\ell-1}
F(\bl_1-a , \ell+1 ; 2\bl_1 ; X/(X-1))$ then \invv\ shows that
\spF\ satisfies \symF{a}.
Using the integral 
\eqn\cnorm{ c_n = {1\over \pi} 
  \int_0^\pi \! \cos^{2n}\! s \, \d s = {(\half)_n \over n!} \, , } 
and the power series expansion of the hypergeometric function in 
\spF\ when $X \to x \cos^2 s$ gives directly 
\eqn\Fz{ 
F^{({1\over 2})}{\!}_{\lambda_1\lambda_2}(x,\bx)  \, \big |_{\lambda_2
= {1\over 2} - b}\limu{\bx\to 0} c_\ell \,  (x \bx)^{\lambda_2} \, x^{\ell} 
F ( \lambda_1+a , \ell + \half ;  2 \lambda_1 ; x ) \, , } 
in accord with the limiting form \Fxsim\  for the choice of 
$c_{\ell}$ given by \cnorm, which is identical to \cell\ for $\vep=\half$.

In an attempt to understand the form of  expressions for more general solutions 
we consider using \relFb
\eqn\lowF{
u^{b+ {3\over 2}}v^{-a-b+1}  \D^{({1\over 2})} u^{b-{1\over 2} }v^{a+b}
\, F^{({1\over 2})}{\!}_{\lambda_1\lambda_2}(a,b)  \, 
= - 2b (\lambda_1 - b) \, 
F^{({1\over 2})}{\!}_{\lambda_1\lambda_2}(a,b-1) \quad \hbox{for} \quad
\lambda_2 = \half  - b \, ,
}
starting from \fint. To apply this  it is sufficient to note that
\eqn\DDF{\eqalign{
& u^{1-p} v^{1-q} \, \D^{({1\over 2})} \big ( u^p v^q \, F(X) \big ) \cr
& {} \sim - \bigg (  q  u \, \Big ( (1-X){\pr \over \pr X}
- p - q - \half \Big ) 
- p v \, \Big ( X {\pr \over \pr X} + p+ q +\half \Big ) -pq  \bigg ) F(X) \cr
&{}  \sim  - \bigg (  q   \, \Big ( X(1-X){\pr \over \pr X} - X 
+ p + \half \Big ) 
- (p+q)  v \, \Big ( X {\pr \over \pr X} + p+ \half \Big ) -(p+q) q\, u  \bigg ) F(X)
\, , 
}
}
using \DFeq\ and
\eqn\fgred{\eqalign{
& \bigg ( u \,  (1-X){\pr \over \pr X} + v \, X {\pr \over \pr X} 
- X (1-X){\pr \over \pr X} + X \bigg ) F(X)\cr
&{}= \half ( 1+u-v) \, F (X) 
+ {\pr \over \pr s}\Big ( {1\over 4} \sin 2s \; (x-\bx) F(X) \Big ) \, .
}
}
Applying \lowF\ to \fint\ then gives
\eqn\Fbm{
2b (\lambda_1 - b) \, F^{({1\over 2})}{\!}_{\lambda_1\lambda_2}(a,b-1)
\big |_{\lambda_2 = {1\over 2} - b}   = 
{1\over \pi} \int_0^\pi \! \d s \;  
\Big ( { u \over X} \Big )^{{1\over 2} - b} \, {\tilde f}(X) \, ,
}
for
\eqn\newf{\eqalign{
{\tilde f}(X) =
\bigg ( & (a+b)\, \Big ( X(1-X){\pr \over \pr X} - (b+\half) X
- b \Big ) \cr
\noalign{\vskip -4pt}
&\hskip 0.5cm {} - (a-b)v\,  \Big ( X {\pr \over \pr X} - b \Big )
- (a^2-b^2) u \bigg ) f(X) \, .
}
}
With $f(X)$ given by \Solfs\ and using \gpb\ we may then obtain directly
from \newf
\eqn\solfn{ \eqalign{
{\tilde f}(X) = & (a+b)(\lambda_1 - a) \, 
g_{\lambda_1}(a-1,b+\half;X) \cr
&{} - (a-b)(\lambda_1 + a)v \, g_{\lambda_1}(a+1,b+\half;X) \cr
&{} + (a^2 - b^2)(1+v-u) \, g_{\lambda_1}(a,b+\half;X) \, .
}
}
For $a=\pm b$ this simplifies to just one term. More generally by repeated
application of $\D_-$ we may express 
$F^{({1\over 2})}{\!}_{\lambda_1\lambda_2}(b,b-n)$
solely in terms of $g_{\lambda_1}(b-n,b+\half;X)$ but this is just equivalent
to \spF\ for $a=b-n$.

\newsec{Discussion}

Explicit expressions for conformal partial waves have proved quite useful
in recent years in attempts to revive the bootstrap approach to conformal
field theories. It is clear that identifying the sum over conformal
blocks in the $s$ and $t$ channels of particular four point functions can
lead to non trivial constraints \refs{\Rychk}, which may be of 
phenomenological interest. On a theoretical level it would be very nice
to apply such ideas in three dimensions, where there is a plethora of
conformal field theories. This paper was begun in an attempt to find an
tractable expression for three dimensional conformal waves which could be
used for similar bootstrap calculations.

In four dimensions there
is a very close connection between the solutions of superconformal Ward
identities and the explicit expressions for conformal partial waves
\refs{\NO} which may be expected to generalise to other dimensions \refs{\ASok}.
In particular the conformal partial waves for leading twist have a special
form and are related to the contributions of semi-short operators in the
operator product expansion. In this paper we have shown how leading twist
conformal partial waves are simplified by being directly related 
to solutions of $\D^{(\vep)} F = 0$ which is just a restriction
of massless wave equation.
The operator  $\D^{(\vep)}$ also appeared in the discussion in  \refs{\ASok}.
This observation demonstrates that the form for conformal partial waves
must be very different in even and odd dimensions. Although we have found
and expression for leading twist in three dimensions there is no straightforward
extension to the arbitrary twist. In even dimensions it is clear, at least in
retrospect, that the construction is very simple starting in two dimensions
where there is, using $x,\bx$, separation of variables and then using
the dimension raising operator $\E_+$. Although the various recurrence
relations obtained in section 3 relate different conformal partial waves
they do not allow expressions for the general case to be obtained from 
leading twist. A natural guess is to suppose that the three dimensional
conformal partial waves can be given by a similar integral expression as in
the leading twist case but involving products of functions $g_p,g_q$,
with different arguments, but a precise form has eluded us so far.

\bigskip

\noindent{\bigbf Acknowledgements}

FAD would like to thank DAMTP Cambridge for hospitality when
much of this work was carried out. We would also like to thank
Jo\~ao Penedones for helpful correspondence.

\vfil\eject

\appendix{A}{Two dimensional conformal integrals}

Although the Symanzik formula \Syman\ provides a general form for a
certain class of conformal integrals in general dimensions in two
dimensions using complex variables there are further extensions. We
consider here integrals of the form
\eqn\twoint{
I_n = {1\over \pi} \int \d^2 x \; f_n(z) \, {\bar f}_n (\bz)  \, , \quad
f_n(z) = \prod_{i=1}^n {1\over (z-z_i)^{\, h_i}} \, , \
{\bar f}_n(\bz) = \prod_{i=1}^n {1\over (\bz-\bz_i)^{\, \bh_i}}  \, ,
}
where
\eqn\hhcon{
{\ts \sum_{i=1}^n} h_i = {\ts \sum_{i=1}^n} \bh_i = 2 \, , \quad
h_i - \bh_i \in {\Bbb Z} \, .
}
The first condition is necessary for $I_n$ to transform covariantly
under conformal transformations, $z \to (a z+ b)/(cz + d)$, 
$\bz \to ({\bar a} \bz+ {\bar b})/({\bar c} \bz + {\bar d})$, and the
second for $I_n$ to be single valued. The integral in \twoint\ reduces
to the form discussed by Symanzik when $h_i = \bh_i$. Convergence
of the integral requires $h_i + \bh_i <2$ for all $i$ although $I_n$ may
be extended by analytic continuation.

When $n=2$
\eqn\Itwo{
I_2 = K_{12} (-1)^{h_1 - \bh_1} \, \pi \delta^2(x_1-x_2) \, ,
}
for
\eqn\Ktwo{
K_{12} = {\Gamma(1-h_1) \, \Gamma(1-h_2) \over
{{\Gamma(\bh_1) \, \Gamma(\bh_2)}} } =
{\Gamma(1-\bh_1) \, \Gamma(1-\bh_2)  \over
\Gamma(h_1) \, \Gamma(h_2) } \, ,
}
with equality of the two forms following from \hhcon\ noting that
$\Gamma(h)\Gamma(1-h)= \pi /\sin \pi h$.

Here we show how evaluation of $I_n$ for $n>2$ can be reduced to solving 
differential equations. To this end we note that
\eqn\difff{
\bigg ( {\pr^2 \over \pr z^2} + {\pr \over \pr z} \sum_{i=2}^n
{h_i \over z- z_i} \bigg ) f_n(z) =
\bigg ( {\pr^2 \over \pr z_1{\!}^2} + \sum_{i=2}^n {1\over z_{1i}} \Big (
h_i {\pr \over \pr z_1} - h_1 {\pr \over \pr z_i} \Big ) \bigg ) f_n(z) \, ,
} 
so that $I_n$ satisfies the second order equation
\eqn\diffI{
\bigg ( {\pr^2 \over \pr z_1{\!}^2} + \sum_{i=2}^n {1\over z_{1i}} \Big (
h_i {\pr \over \pr z_1} - h_1 {\pr \over \pr z_i} \Big ) \bigg ) I_n =0 \, .
}
There is of course a corresponding equation for $z_i \to \bz_i, \, 
h_i \to \bh_i$.

For $n=3$ the form of $I_n$ for non coincident $x_i$ is dictated just 
by conformal invariance
\eqn\Ithree{
I_3 = K_{123} \, z_{12}{\!}^{h_3-1} z_{23}{\!}^{h_1-1} z_{31}{\!}^{h_2-1} \,
\bz_{12}{\!}^{\bh_3-1} \bz_{23}{\!}^{\bh_1-1} \bz_{31}{\!}^{\bh_2-1} \, ,
}
where the overall constant can be obtained by extending the result
for $h_i = \bh_i$
\eqn\Kthree{
K_{123} = {\Gamma(1-h_1) \, \Gamma(1-h_2) \, \Gamma(1-h_3) \over
{{\Gamma(\bh_1) \, \Gamma(\bh_2) \, \Gamma(\bh_3)}} } = 
{\Gamma(1-\bh_1) \, \Gamma(1-\bh_2) \, \Gamma(1-\bh_3) \over
\Gamma(h_1) \, \Gamma(h_2) \, \Gamma(h_3) } \, .
}
It is easy to verify that \Kthree\ satisfies \diffI\ subject to \hhcon,
which is also necessary to ensure the required symmetries under permutations
of $\{z_i \}$ or $\{\bz_i \}$.

For $n=4$ conformal invariance shows that the result must have the form
\eqn\Ifour{
I_4 = z_{12}{\!}^{h_3+h_4-1} z_{23}{\!}^{h_1+h_4-1} z_{31}{\!}^{h_2-1} 
z_{24}{\!}^{-h_4}\,
\bz_{12}{\!}^{\bh_3+\bh_4-1} \bz_{23}{\!}^{\bh_1+\bh_4-1} \bz_{31}{\!}^{\bh_2-1}
\bz_{24}{\!}^{-\bh_4}\, \I(x,\bx) \, ,
}
for $\I(x,\bx)$ an undetermined functions of the invariants $x,\bx$ defined
here by \xzz. However imposing \diffI\ leads to the differential equation
\eqn\diffIfour{
\bigg (x(1-x) {\pr^2 \over \pr x^2} + \big ( h_3+h_4 + ( h_2 - h_4 - 2) x 
\big ) {\pr \over \pr x} + h_4 (h_2-1) \bigg ) \I(x,\bx)=0  \, ,
}
which has a simple hypergeometric form. The relevant solutions of
\diffIfour\ and its conjugate can be expressed as
\eqn\Iresult{\eqalign{
\I(x,\bx) = {}& K_4 \, F(1-h_2, h_4; h_3+h_4 ; x) \, 
F(1-\bh_2, \bh_4; \bh_3+\bh_4 ; \bx) \cr
&{} + \bK_4 \, (-1)^{h_1+h_4-\bh_1-\bh_4}\,
x^{h_1+h_2-1} F(1-h_3, h_1; h_1+h_2 ; x) \cr
\noalign{\vskip -4pt}
&{} \hskip 3.3cm {}\times
\bx^{\bh_1+\bh_2-1} F(1-\bh_3, \bh_1; \bh_1+\bh_2 ; \bx)  \, .
}
}

Although solving the differential equation allows a more general form
than \Iresult\ the result given is required by the symmetry relations
\eqnn\Isymm
$$\eqalignno{
\I(x,\bx) = {}& (1-x)^{-h_4}(1-\bx)^{-\bh_4} \,
\I(x',\bx')\, \big |_{h_1 \leftrightarrow h_2,\bh_1 \leftrightarrow \bh_2} \, , 
\qquad x' = {x \over x-1} \, , \ \bx' = {\bx \over \bx -1} \, ,\cr
\I(x,\bx) = {}& (1-x)^{h_2 -1}(1-\bx)^{\bh_2-1}
\I(x',\bx')\, \big |_{h_3 \leftrightarrow h_4,\bh_3 \leftrightarrow \bh_4}\, ,
& \Isymm
}
$$
and
\eqn\Isym{
\I(x,\bx) = (-1)^{h_4 - \bh_4} \, \I(1-x,1-\bx) \,
\big |_{h_1 \leftrightarrow h_3,\bh_1 \leftrightarrow \bh_3} \, .
}
All other symmetry relations can be obtained by combining \Isymm\ and \Isym.
The relations in \Isymm\ follow easily from standard hypergeometric identities
which give $ F(a,b;c ; x) = (1-x)^{-b}  F(c-a,b;c ; x')
= (1-x)^{-a}  F(a,c-b; c ; x')$ if both $K_4, \bK_4$ are invariant under
$1\leftrightarrow 2$ and also $3\leftrightarrow 4$.
To satisfy \Isym\ we need to use the further hypergeometric identities
\eqn\hypersy{\eqalign{
F(1-h_2,&  h_4; h_1+h_4 ; 1-x) \cr
= {}& {\Gamma(h_1+h_4)\Gamma(h_1+h_2-1) \over
\Gamma(1-h_3)\Gamma(h_1)} \, F(1-h_2, h_4; h_3+h_4 ; x) \cr
&{} + {\Gamma(h_1+h_4)\Gamma(h_3+h_4-1) \over
\Gamma(1-h_2)\Gamma(h_4)} \, x^{h_1+h_2-1} F(1-h_3, h_1; h_1+h_2 ; x) \, ,\cr
(1-x)^{h_2+h_3-1}& F(1-h_1, h_3; h_2+h_3 ; 1-x) \cr
= {}& {\Gamma(h_2+h_3)\Gamma(h_1+h_2-1) \over
\Gamma(1-h_4)\Gamma(h_2)} \, F(1-h_2, h_4; h_3+h_4 ; x) \cr
&{} + {\Gamma(h_2+h_3)\Gamma(h_3+h_4-1) \over
\Gamma(1-h_1)\Gamma(h_3)} \, x^{h_1+h_2-1} F(1-h_3, h_1; h_1+h_2 ; x) \, .
}
}
With the aid of \hypersy\ and the associated conjugates we may determine
\eqn\KK{\eqalign{
K_4 = {}& {\Gamma(1-h_1) \, \Gamma(1-h_2) \, \Gamma(h_1+h_2-1) \over
{{\Gamma(\bh_1) \, \Gamma(\bh_2) \, \Gamma(2-\bh_1-\bh_2)}} } \, , \cr
\bK_4 = {}&  {\Gamma(1-h_3) \, \Gamma(1-h_4) \, \Gamma(h_3+h_4-1) \over
{{\Gamma(\bh_3) \, \Gamma(\bh_4) \, \Gamma(2-\bh_3-\bh_4)}} } \, .
}
}
As for $K_{12},K_{123}$ these are symmetric under 
$h_i \leftrightarrow \bh_i$. The overall scale is fixed since we require
$K_4|_{h_4=\bh_4=0}=K_{123}$, $\bK_4|_{h_4=\bh_4=0}= 0$.
To verify that \Isym\ is satisfied it is necessary to repeatedly
use $\Gamma(h)\Gamma(1-h) = \pi / \sin \pi h$ and identities
such as
\eqn\sinid{\eqalign{
& {\sin \pi h_1 \; \sin \pi \bh_3  \over \sin \pi (\bh_3+\bh_4) \;
\sin \pi (h_1 + h_4)} \cr
&{} + (-1)^{h_3+h_4-\bh_3-\bh_4} \, 
{\sin \pi h_2 \; \sin \pi \bh_4  \over \sin \pi (\bh_3+\bh_4) \;
\sin \pi (h_2 + h_3)} = (-1)^{h_4 -\bh_4} \, ,
}
}
which depend on \hhcon.

An alternative more direct evaluation may be obtained by introducing
the conformal covariant differential operator
\eqn\Dhh{
\D^{(h,h')}_{z,z'} = \sum_{n=0}^\infty {(-1)^n \over n!} \, 
{(h')_n \over (h+h')_n} \, (z-z')^n {\pr^n \over \pr z^n} \, ,
}
which satisfies
\eqn\Dzw{
\D^{(h,h')}_{z,z'} {1\over (w-z)^{h+h'}} =
{1\over (w-z)^h \, (w-z')^{h'}} \, .
}
Using this operator we may write
\eqn\DDzz{\eqalign{
& {1\over (z-z_1)^{h_1} (z-z_2)^{h_2}} \, 
{1\over (\bz-\bz_1)^{\bh_1} (\bz-\bz_2)^{\bh_2}} \cr
&{} = \D^{(h_1,h_2)}_{z_1,z_2}\D^{(\bh_1,\bh_2)}_{\bz_1,\bz_2}\,
{1\over (z- z_1)^{h_1+h_2} \, (\bz- \bz_1)^{\bh_1+\bh_2}} \cr
&\quad {}+ K_{4}(-1)^{h_2-\bh_2}
 \, z_{12}{}^{1-h_1-h_2} \bz_{12}{}^{1-\bh_1-\bh_2} \,
\D^{(1-h_2,1-h_1)}_{z_1,z_2}\D^{(1-\bh_2,1-\bh_1)}_{\bz_1,\bz_2}
\, \pi \delta^2(x-x_1) \, , 
} 
}
with $K_4$ as in \KK. The coefficient is dictated by the requirement
that using \DDzz\ in the integral expression for $I_3$ gives the
result \Ithree.

Using \DDzz\ in $I_4$ gives
\eqn\IfDD{\eqalign{
I_4 ={}& K_{4}(-1)^{h_2-\bh_2}
 \, z_{12}{}^{1-h_1-h_2} \bz_{12}{}^{1-\bh_1-\bh_2} \cr
\noalign{\vskip -4pt}
&\quad {}\times
\D^{(1-h_2,1-h_1)}_{z_1,z_2} z_{13}{\!}^{-h_3} z_{14}{\!}^{-h_4} \,
\D^{(1-\bh_2,1-\bh_1)}_{\bz_1,\bz_2}\bz_{13}{\!}^{-\bh_3} \bz_{14}{\!}^{-\bh_4}
\cr
&{}+  \bK_{4}(-1)^{h_3-\bh_3}
 \, z_{34}{}^{1-h_3-h_4} \bz_{34}{}^{1-\bh_3-\bh_4} \cr
\noalign{\vskip -4pt}
&\quad {}\times
\D^{(h_1,h_2)}_{z_1,z_2} z_{13}{\!}^{h_4-1} z_{14}{\!}^{h_3-1} \,
\D^{(\bh_1,\bh_2)}_{\bz_1,\bz_2}\bz_{13}{\!}^{\bh_4-1} \bz_{14}{\!}^{\bh_4-1}\,,
}
}
where agreement with \Ifour\ and \Iresult\ follows from
\eqn\Dzz{\eqalign{
\D^{(h_1,h_2)}_{z_1,z_2} z_{13}{\!}^{h_4-1} z_{14}{\!}^{h_3-1} 
=  {}& z_{13}{\!}^{-h_1} z_{24}{\!}^{h_3-1} z_{23}{}^{h_1+h_4-1} \,
F(1-h_3,h_1;h_1+h_2;x) \, , \cr
\D^{(1-h_2,1-h_1)}_{z_1,z_2} z_{13}{\!}^{-h_3} z_{14}{\!}^{-h_4} = {}&
 z_{13}{\!}^{h_2-1} z_{24}{\!}^{- h_4} z_{23}{}^{h_1+h_4-1} \,
F(1-h_2,h_4;h_3+h_4;x) \, ,
}
}
together with their conjugates.

The verification of \Dzz\ follows from the integral representations
\eqn\demoD{\eqalign{
& \D^{(h_1,h_2)}_{z_1,z_2} z_{13}{\!}^{h_4-1} z_{14}{\!}^{h_3-1} \cr
&{}= {\Gamma(h_1+h_2) \over \Gamma(1-h_3)\,\Gamma(1-h_4)} \int_0^1 \!\!\d 
\alpha \;  {\alpha^{-h_4}(1-\alpha)^{-h_3} \over 
( \alpha z_{32} + (1-\alpha) z_{42})^{h_2} \,
( \alpha z_{31} + (1-\alpha) z_{41})^{h_1}} \cr
&{} =  z_{13}{\!}^{-h_1} z_{24}{\!}^{h_3-1} z_{23}{}^{h_1+h_4-1} \,
{\Gamma(h_1+h_2) \over \Gamma(1-h_3)\,\Gamma(1-h_4)\,\Gamma(h_1)\,
\Gamma(h_2)} \cr
\noalign{\vskip -4pt} &\quad {}\times
{1\over 2\pi i} \int_{-i\infty}^{i\infty} \!\! \d s \;
\Gamma(-s)\,\Gamma(h_1+s)\,\Gamma(1-h_3+s)\,\Gamma(1-h_1-h_4 -s) \,
\Big ( {z_{14}z_{23}\over z_{13}z_{24}} \Big )^{\! s} \, .
}
}

\appendix{B}{Compact case}

In this appendix we demonstrate how the recurrence relations for
conformal partial waves are closely related to those for symmetric
two variable Jacobi polynomials associated with the root system
$BC_2$ described in \Koop\ which contains related results.

For the elementary single variable polynomials we first consider the
differential operator
\eqn\hyperD{ 
\tD_x
= x(1-x) {\d^2 \over \d x^2} + \big ( 1 +\beta - (2+\gamma) x \big ) 
{\d \over \d x} \, , }
and then Jacobi polynomials of degree $n$ may be defined by the eigenvalue 
equation
\eqn\eigJ{
-\tD_x \, P_n^{(\alpha,\beta)}(y) = n (n+\gamma+1) \, 
P_n^{(\alpha,\beta)}(y) \, , \qquad y=2x-1 \, , 
}
where we require
\eqn\gabc{
\gamma =  \alpha + \beta \, .
}

It is convenient here to adopt a non standard normalisation for Jacobi
 polynomials $R_n^{(\alpha,\beta)} (y)$ such that $R_n^{(\alpha,\beta)}(1)=1$.
Recursion relations may be obtained by following 
the same procedure as in section 3. Defining
${\tilde f}_0 = x -\half , \,  {\tilde f}_1 = x(1-x) {\d \over \d x}$
then the commutators 
\eqn\Ccom{
[ \tD_x , {\tilde f}_0 ] = 2 {\tilde f}_1 
- (2+\gamma) {\tilde f}_0 + \half(\beta-\alpha) \, , \quad
[ \tD_x , {\tilde f}_1 ] = - 2 {\tilde f}_0 \tD_x + \gamma {\tilde f}_1 \, ,
}
require
\eqn\recR{\eqalign{
{\tilde f}_0 \, R_n^{(\alpha,\beta)} = {}& a \, R_{n-1}^{(\alpha,\beta)} + 
b \, R_{n+1}^{(\alpha,\beta)}
+ {\half \gamma(\beta-\alpha) \over (2n+\gamma)(2n+\gamma+2)} \, 
R_n^{(\alpha,\beta)} \, ,\cr
{\tilde f}_1 \, R_n^{(\alpha,\beta)} = {}& (n+\gamma+1)a \, 
R_{n-1}^{(\alpha,\beta)} -n b \, R_{n+1}^{(\alpha,\beta)}
- {n(n+ \gamma+1)(\beta-\alpha) \over (2n+\gamma)(2n+\gamma+2)} \, 
R_n^{(\alpha,\beta)} \, .
}
}
Imposing the normalisation conditions, so that 
${\tilde f}_0 \, R_n^{(\alpha,\beta)}\big |_{x=1}= \half, \,
{\tilde f}_1 \, R_n^{(\alpha,\beta)}\big |_{x=1}= 0$, determines
\eqn\resab{
a = {n(n+\beta)\over (2n+\gamma)(2n+\gamma+1)} \, , \qquad
b = {(n+\gamma+1)(n+\alpha+1)\over (2n+\gamma)(2n+\gamma+1)} \, .
}

The corresponding two variable operator is then constructed using
$\tD$ defined in \hyperD
\eqn\defDJ{
\tDelta^{(\vep)}  =  -\tD_x -  \tD_\bx 
-  2\vep \, {1 \over x-\bx} \Big (x (1-x) {\pr \over \pr x} - \bx (1- \bx) 
{\pr \over \pr \bx} \Big ) \, , 
}
and the eigenvalue equations becomes
\eqn\eigDJ{
\tDelta^{(\vep)}  R_{nm}^{(\alpha,\beta,\vep)} = {\tilde c}_{nm} \,
R_{nm}^{(\alpha,\beta,\vep)} \, ,
}
where $ R_{nm}^{(\alpha,\beta,\vep)}(y,{\bar y})$ are symmetric 
polynomials, $y=2x-1,{\bar y} = 2\bx -1$, with the convenient normalisation
\eqn\normR{
R_{nm}^{(\alpha,\beta,\vep)} (1,1) = 1 \, .
}
For polynomial solutions such that
\eqn\limR{
R_{nm}^{(\alpha,\beta,\vep)} (y,{\bar y}) \limu{\bx \to \infty} k_{nm} \, 
\bx^n P_m^{(\alpha,\beta)}(y)   \, , \qquad n\ge m \, ,
}
then \eigDJ\ determines
\eqn\cJ{
{\tilde c}_{nm} = n (n+ 1 + \gamma+2\vep) + m(m+1 +\gamma) \, , \quad 
m,n = 0, 1 , \dots \, , \ m\le n \, .
}

In a similar fashion to \pid
\eqn\pidJ{
\tDelta^{(\vep)} \,( x-\bx )^{1 - 2\vep} =
( x-\bx )^{1- 2\vep } \big ( \tDelta^{(1-\vep)}
+ (2+\gamma) (1 - 2\vep ) \big ) \, . }
Also for
\eqn\EpdefJ{
{\tilde \E}_+ = - {1\over x-\bx} \, \big ( \tD_x - \tD_\bx \big ) \, ,
}
then analogous to \EpD
\eqn\EpJ{
{\tilde \E}_+ \, \tDelta^{(\vep)} =
\big (  \Delta^{(\vep+1)} + 2+ \gamma + 2\vep \big )\, {\tilde \E}_+ \, .
}
Using \pidJ\ if
\eqn\pEM{
{\tilde \E}_-{\!}^{(\vep+1)} = (x-\bx)^{1-2\vep} {\tilde \E}_+
(x-\bx)^{1+2\vep} =  (x-\bx)^{-2\vep}\big ( \tD_x - \tD_\bx \big )
(x-\bx)^{1+2\vep} \, ,
}
then
\eqn\EmJ{
{\tilde \E}_-{\!}^{(\vep+1)} \Delta^{(\vep+1)} =
\big (  \Delta^{(\vep)} + 2+ \gamma + 2\vep \big )\, 
{\tilde \E}_-{\!}^{(\vep+1)} \, .
}
As a consequence of \EpJ\ and \EmJ
\eqn\rRR{
{\tilde \E}_+ \, R_{n\,m}^{(\alpha,\beta,\vep)} =
a^{(\vep)} \, R_{n-1 \, m}^{(\alpha,\beta,\vep+1)} \, , \qquad
{\tilde \E}_-{\!}^{(\vep+1)}   R_{n \, m}^{(\alpha,\beta,\vep+1)} 
= b^{(\vep)} \,  R_{n+1\, m}^{(\alpha,\beta,\vep)} \, .
}
Requiring \normR\ the coefficient $b^{(\vep)}$ may be directly determined by
\eqn\resb{
b^{(\vep)} = (x-\bx)^{-2\vep}\big ( \tD_x - \tD_\bx \big )
(x-\bx)^{1+2\vep} \big |_{x=\bx=1} =  2 (1+2\vep)(\alpha + \vep + 1 ) \, .
}
To calculate $a^{(\vep)}$ it is sufficient to note that
\eqn\Casz{
\tDelta_4{\!}^{(\vep)} =  {\tilde \E}_-{\!}^{(\vep+1)} {\tilde \E}_+ =
( x-\bx )^{- 2 \vep} \big ( \tD_x - \tD_\bx \big ) \,
( x-\bx )^{2\vep}  \big ( \tD_x - \tD_\bx \big ) \, ,
}
is a fourth order Casimir operator such that
\eqn\CasfJ{
\tDelta_4{\!}^{(\vep)}   R_{nm}^{(\alpha,\beta,\vep)} = {\tilde c}_{4,nm} \,
R_{nm}^{(\alpha,\beta,\vep)} \, ,
}
and hence
\eqn\resa{ 
a^{(\vep)} \, b^{(\vep)} =  {\tilde c}_{4,nm}  \, .
}
Applying \limR\ with \Casz\ dictates
\eqn\eigfJ{
{\tilde c}_{4,nm}  
= (n-m)(n-m+2\vep)(n+m+1+\gamma)(n+m+1+\gamma + 2\vep) \, .
}

Just as $F_{\lambda_1 \lambda_2}(x,\bx)$  determine the conformal partial 
waves $G_\Delta^{(\ell)}$ there are harmonic polynomials 
associated with $R_{nm}^{(\alpha,\beta,\vep)}$ given by
\eqn\YR{
R_{nm}^{(\alpha,\beta,\vep)} (y,{\bar y}) = 
Y_{nm}^{(\alpha, \beta)}(\sigma,\tau)\, ,
\qquad \sigma = x \bx \, , \ \tau=(1-x)(1-\bx) \, ,
}
which play a role \NO\ in the decomposition of correlation functions. The
differential operator acting on $ Y_{nm}^{(\alpha, \beta)}(\sigma,\tau)$ becomes
\eqn\diffY{\eqalign{
{\tilde \D} = {}& - ( 1-\sigma-\tau) \bigg ( {\pr \over \pr \sigma} \sigma
 {\pr \over \pr \sigma} +  {\pr \over \pr \tau} \tau  {\pr \over \pr \tau}
+ \beta \,  {\pr \over \pr \sigma} + \alpha \,  {\pr \over \pr \tau} \bigg )
+ 4 \, \sigma\tau \,  {\pr^2 \over \pr \sigma \pr \tau} \cr
&{} + (d+2\alpha ) \, \sigma {\pr \over \pr \sigma} + (d+2\beta ) \,
 \tau  {\pr \over \pr \tau} \, ,
}
}
so that
\eqn\eigY{
{\tilde \D} \, Y_{nm}^{(\alpha, \beta)}(\sigma,\tau) = {\tilde c}_{nm} \,
 Y_{nm}^{(\alpha, \beta)}(\sigma,\tau)\, .
}
The relevant boundary conditions are then
\eqn\Ybo{
 Y_{nm}^{(\alpha, \beta)}(\sigma,\tau) \propto \sigma^n \bigg ( 1 - 
{\tau \over \sigma} \bigg )^{\! n-m} \, , \qquad \sigma \to \infty \, , \
\tau = {\rm O}(\sigma) \, .
}

It is easy to see that
\eqn\Rzero{
R_{nm}^{(\alpha,\beta,0)} (y,{\bar y}) = \half ( R_{n}^{(\alpha,\beta)} (y)\, 
R_{m}^{(\alpha,\beta)} ({\bar y}) +  R_{n}^{(\alpha,\beta)} ({\bar y}) \,
R_{m}^{(\alpha,\beta)} (y) \big ) \, .
}
Starting from \Rzero\ and using \rRR\ with $a^{(0)}$ given by \resa\ gives
\eqn\Rone{\eqalign{
(n-m+1)(n+m+2 + \gamma) \,& R_{nm}^{(\alpha,\beta,1)} (y,{\bar y}) \cr
= {}& 2(\alpha+1) \, {R_{n+1}^{(\alpha,\beta)} (y)
R_{m}^{(\alpha,\beta)} ({\bar y}) -  R_{n+1}^{(\alpha,\beta)} ({\bar y})
R_{m}^{(\alpha,\beta)} (y) \over y - {\bar y}}  \, .
}
}
The results \Rzero\ and \Rone\ correspond to \Fzero\ and \two\ for conformal
partial waves.

To establish further recurrence relations for $ R_{nm}^{(\alpha,\beta,\vep)} $ 
following the same method as in the non compact case we define
\eqn\deftF{\eqalign{
\tF_0 = {}& x + \bx - 1 \, ,  \qquad
\tF_1 = x(1-x){\pr \over \pr x} + \bx (1-\bx){\pr \over \pr \bx}  \, , \cr
\tF_2 = {}& ( x-\bx )  \, \big ( \tD_x - \tD_\bx \big ) \, , \cr
\tF_3 = {}& \bigg ( x (1-x){\pr \over \pr x} -  \bx(1-\bx){\pr \over \pr \bx} 
+ 2\vep \,  {x(1-x)+\bx(1-\bx) \over x- \bx} \bigg )  
\big ( \tD_x - \tD_\bx \big ) \, , 
}
}
and then, similarly to \comD,
\eqn\comDJ{\eqalign{
\big [ \tDelta^{(\vep)} , \tF_0 \big ] ={}& - 2 \, \tF_1 + (2+\gamma+2\vep ) \, 
\tF_0 + \alpha - \beta\, ,  \cr
\qquad \big [ \tDelta^{(\vep)} , \tF_1 \big ] = {}&  \tF_2 - \gamma \, \tF_1  - 
\tF_0\,  \tDelta^{(\vep)}   \, , \cr
\big [ \tDelta^{(\vep)} , \tF_2 \big ] = {}& - 2\, \tF_3
+ (2 + \gamma -2 \vep)\, \tF_2  \, , \cr
\big [ \tDelta^{(\vep)} , \tF_3 \big ] = {}&  \tF_0 \, 
\tDelta_4{\!}^{(\vep)} - \gamma \, \tF_3{\!}^{(\vep)}
- \tF_2 \big (\tDelta^{(\vep)} + 2\vep(1+\gamma )  \big ) \, .
}
}

Just as in the non compact case in section 4 the algebra \comDJ\ dictates
that
\eqn\FRR{
\tF_i R_{nm}^{(\alpha,\beta,\vep)}= a_i \, 
R_{n-1\, m}^{(\alpha,\beta,\vep)} + b_i \, R_{n+1\, m}^{(\alpha,\beta,\vep)}
+ c_i \, R_{n\, m-1}^{(\alpha,\beta,\vep)}
+ d_i \, R_{n\, m+1}^{(\alpha,\beta,\vep)} 
+ u_i \, R_{nm}^{(\alpha,\beta,\vep)} \, ,
}
where
\eqnn\abcdu$$\eqalignno{
a_1 = {}& ( n + 1+ \gamma + 2 \vep) a_0 \, , \quad 
b_1 = - n\, b_0 \, , \quad c_1= (m + 1 + \gamma +  \vep ) c_0 \, , \quad
d_0 = - (m-\vep) d_0 \, , \cr
a_2 ={}& - (n-m+2\vep)(n+m + 1 + \gamma + 2 \vep) a_0 \, , \quad
b_2 = -(n-m)(n+m + 1 + \gamma )b_0 \, , \cr
c_2 ={}& (n-m) (n+m + 1 + \gamma + 2 \vep) c_0 \, , \quad
d_2 =(n-m+2\vep) (n+m + 1 + \gamma  ) d_0 \, , \cr
a_3 = {}& - (n+ 1 + \gamma) (n-m+2\vep)(n+m + 1 + \gamma + 2 \vep) a_0 \, , \cr
b_3 = {}&  (n+2\vep) (n-m)(n+m + 1 + \gamma)b_0 \, , \cr
c_3 = {}& (m+ 1 + \gamma-\vep) (n-m) (n+m + 1 + \gamma + 2 \vep) c_0 \, , \cr
d_3 = {}& - (m+\vep)(n-m+2\vep) (n+m + 1 + \gamma) d_0 \, , & \abcdu
}
$$
as well as
\eqn\ures{\eqalign{
u_0 = {}& - {(\alpha-\beta) \gamma \, \big (2 {\tilde c}_{n,m} + 
(\gamma+2)(\gamma + 2 \vep) \big ) \over
(2n+2 + \gamma + 2 \vep) (2n + \gamma + 2 \vep)(2m +2 + \gamma)
(2m + \gamma) } \, , \cr
u_2 = {}& - {2(\alpha-\beta) \gamma \, {\tilde c}_{4,nm} \over
(2n+2 + \gamma + 2 \vep) (2n + \gamma + 2 \vep)(2m +2 + \gamma)
(2m + \gamma) } \, , \cr
u_1 ={}& \half (2+ \gamma + 2 \vep)u_0 + \half(\alpha-\beta) \, , \quad
u_3 = \half (2+ \gamma - 2 \vep)u_2 \, .
}
}
The coefficients $a_0,b_0,c_0,d_0$ are determined by applying the
normalisation conditions at $x=\bx=1$ in \FRR\ which imply
$a_i+b_i+c_i+d_i+u_i = 1, \, i=0, \, 0 , \, i=1,2,3$. Hence
\eqn\abcdz{\eqalign{
a_0 = {}&  {(n+\vep)(n+\beta+\vep)(n-m)(n+m+1+\gamma) \over
(2n+ \gamma + 2 \vep)(2n+1+\gamma+2\vep)(n-m+\vep)(n+m+1+\gamma+\vep)}\, , \cr
b_0 = {}&  {(n+1+\gamma+\vep)(n+1+\alpha+\vep)(n-m+2\vep)(n+m+1+\gamma+2\vep) 
\over (2n+ 1+ \gamma + 2 \vep)(2n+2+\gamma+2\vep)
(n-m+\vep)(n+m+1+\gamma+\vep)}\, , \cr
c_0 = {}&  {m(m+\beta)(n-m+2\vep)(n+m+1+\gamma) \over
(2m + \gamma)(2m+1+\gamma)(n-m+\vep)(n+m+1+\gamma+\vep)}\, , \cr
d_0 = {}&  {(m+1+\gamma)(m+1+\alpha)(n-m)(n+m+1+\gamma+2\vep) \over
(2m +1+ \gamma)(2m+2+\gamma)(n-m+\vep)(n+m+1+\gamma+\vep)}\, . \cr
}
}

The results for $a_i,b_i,c_i,d_i$ are related for each $i$ 
as a consequence of the symmetry relations
\eqn\extR{
R_{nm}^{(\alpha,\beta,\vep)}  = 
R_{-n -1 - \gamma - 2 \vep \, - m - 1 - \gamma}^{(\alpha,\beta,\vep)} 
= R_{m - \vep \, n + \vep }^{(\alpha,\beta,\vep)} \, ,
}
From \EpdefJ\ and  \deftF\ ${\tilde \E}_+ = -(x-\bx)^{-2}\tF_2$ so that
\rRR\ and \FRR\ for $i=2$ with
\eqn\abcdt{\eqalign{
a_2 = {}& - {(n+\vep)(n+\beta+\vep)\, \,{\tilde c}_{4,nm} \over
(2n+ \gamma + 2 \vep)(2n+1+\gamma+2\vep)(n-m+\vep)(n+m+1+\gamma+\vep)}\, , \cr
b_2 = {}& - {(n+1+\gamma+\vep)(n+1+\alpha+\vep)\, {\tilde c}_{4,nm}
\over (2n+ 1+ \gamma + 2 \vep)(2n+2+\gamma+2\vep)
(n-m+\vep)(n+m+1+\gamma+\vep)}\, , \cr
c_2 = {}&  {m(m+\beta)\, {\tilde c}_{4,nm} \over
(2m + \gamma)(2m+1+\gamma)(n-m+\vep)(n+m+1+\gamma+\vep)}\, , \cr
d_2 = {}&  {(m+1+\gamma)(m+1+\alpha)\, {\tilde c}_{4,nm} \over
(2m +1+ \gamma)(2m+2+\gamma)(n-m+\vep)(n+m+1+\gamma+\vep)}\, , \cr
}
}
determine $ R_{nm}^{(\alpha,\beta,\vep+1)}$ in 
terms of $ R_{n'm'}^{(\alpha,\beta,\vep)}$ for $n'=n,n+2,m'=m$ and
$n'=n+1, m'= m, m\pm 1$,  in agreement with the result in \Vretare.

\appendix{C}{Finiteness for $d=6$}

The linear combination of $\F^-{\!\!}_{pq}$'s appearing  in \four\  is 
determined in order to ensure the result is finite as $\bx \to x$. To 
verify this we consider the Wronskian
\eqn\Wron{
W_{pq}(x) = g_p{\!}'(x) \, g_q(x) - g_q{\!}'(x) \, g_p(x) \, .
}
The defining equation  \defg\ then gives
\eqn\Weq{
{\d  \over \d x} \Big ( (1-x)^{a+b-1} \, W_{pq}(x) \Big ) =
(p-q)(p+q-1) \,   x^{-1}(1-x)^{a+b-1} \, g_p(x) g_q(x) \, .
}
Hence
\eqn\WWeq{\eqalign{
& {\d  \over \d x} \Big ( (1-x)^{a+b-1} \big ( (p-q)\, W_{p+1\,q-1}
- (p-q+2) \, W_{pq}\big ) \Big )  \cr
& {} =
(p-q)(p-q+2) (p+q-1) \,   x^{-1}(1-x)^{a+b-1} \, \big (
g_{p+1} g_{q-1} - g_p\,  g_q\big ) \, .
}
}
Using the identity \idgp\ in $\big (g_{p+1}(x)/x \big ) g_q(x) = g_{p+1}(x)
\big ( g_q(x) /x\big )$ gives
\eqn\ggg{
g_{p+1} g_{q-1} - g_p\,  g_q = ( \alpha_{p+1} - \alpha_q)\,  g_{p+1} \, g_q
+ \beta_{p+1} \, g_{p+2}  g_q - \beta_q \, g_{p+1} g_{q+1} \, .
} 
As a  consequence of \WWeq\ this leads to the relation
\eqn\wrel{\eqalign{
(p-q)& \, W_{p+1\,q-1} - (p-q+2) \, W_{pq} \cr
={}&  {p+q -1\over p+q} \, {(p-q)(p-q+2) \over p-q+1} \, 
( \alpha_{p+1} - \alpha_q) W_{p+1\, q} \cr
&{}+ {p+q-1\over p+q+1} \, \big ( 
(p-q)\beta_{p+1} \, W_{p+2\,q} - (p-q+2)\beta_q \, W_{p+1\, q+1} \big ) \, .
}
}
Since $\F^-{\!}_{pq}(x,\bx) = (x-\bx) W_{pq}(t) + {\rm O}((x-\bx)^3)$, 
$t=\half(x+\bx)$, the coefficients in \four\ are exactly of the form 
necessary according to \wrel\ to cancel the singular terms as $x\to \bx$.

\appendix{D}{Free Theory and Leading Twist}

In four dimensions the conformal partial wave expansion of
the four point function obtained from elementary scalar fields,
with $\Delta=1$, has only contributions from twist two operators.
Here we extend this to other cases. With $\D^{(\vep)}$ given
by \defDD\ it is easy to see that $\D^{(\vep)} u^\alpha v^\alpha =0$
has solutions only for $(\alpha,\beta)=(0,0),(-\vep,0),(0, -\vep)$.
Assuming all operators in the four point function have
the same scale dimension, $\Delta_1=\Delta_2=\Delta_3=\Delta_4=  \vep$
so that $a=b=0$, then we identify  the four point function for 
elementary scalars as satisfying $\D^{(\vep)} u^{-\vep} F(u,v) = 0$
so that it has the general form
\eqn\Ffree{
F(u,v) = 1 + C \, u^{\vep} + C' \, \Big ( {u\over v} \Big )^{\vep} \, ,
}
where $1$ corresponds to the identity  in the operator product expansion
and the two point function has been normalised to 1. Crossing symmetry
requires also $C=C'=1$ but we relax that here for generality.

As a consequence of \DFh, with $b=0$, $N=1$, the conformal partial wave
expansion is restricted to be 
\eqn\freeW{
F(u,v) = \sum_{\lambda_1} a_{\lambda_1} \, 
F_{\lambda_1 \vep} (x,\bx) \, .
}
To determine the coefficients $a_{\lambda_1}$ it is sufficient to consider
only $x=\bx$ as the solutions of $\D^{(\vep)} f(x,\bx) = 0 $ are determined 
in terms of $f(x,x)$. From \Ffree\ we have
\eqn\Ffree{
F(u,v)\big |_{x=\bx} = 1 + C \, x^{2\vep} + C' \, 
\Big ( {x\over 1-x} \Big )^{2\vep} \, ,
}

For $\vep =\half$ then from \thF\ and \hFf\ with \fint\ and \Solfs\ the 
partial wave expansion \freeW\ requires
\eqn\threew{
 x^{-{1\over 2}}+ C \, x^{{1\over 2}} + C'   {x^{{1\over 2}}\over 1-x}  = 
\sum_{\lambda_1} a_{
\lambda_1} \, g_{\lambda_1}(0,\half; x) \, .
}
This has a solution 
\eqn\apt{
a_{- {1\over 2}} = 1 \, , \quad a_{{1\over 2}} = C+C' \, \quad
a_{p+{1\over 2}} = \big ( C(-1)^p + C' \big ) \, 2^{-(2p-1)} \quad
\hbox{for} \quad p=1,2,\dots \,.
} 
Note that in this case $F_{-{1\over 2}\, {1\over2}} = F_{0\, 0} =1$.

When $\vep=1$ we may use \ftwo, with $b=0$, so that \freeW\ with
\Ffree\ become
\eqn\twoexp{
F(u,v)\big |_{x=\bx} = x^2 \sum_{\lambda_1} {1\over \lambda_1}a_{\lambda_1} 
g_{\lambda_1}{\!}'(x) \quad \Rightarrow \quad
 -{1\over x}  + C \, x + C' \,  {1\over 1-x} + a = \sum_{\lambda_1}  {1\over \lambda_1}
a_{\lambda_1} \,  g_{\lambda_1}(x) \, ,
}
for some arbitrary constant $a$. Matching the power expansions on both
sides gives
\eqn\twores{
a_{-1} = 1 \, , \qquad a_{p+1} = \big ( C(-1)^p + C' \big ) \, {p!\, (p+1)!\over (2p)!}\, ,
\quad p=0,1,\dots \, ,
}
agreeing with the result in \DO\ since for $\vep =1$ $F_{-1 1} = F_{00} =1$.

In the six dimensional case, $\vep=2$, we may use from \ffour\ and \fsol,
with $a,b=0$,
$(\lambda_1+1)\lambda_1(\lambda_1-1) F_{\lambda_1 2}(x,x)
= x^4 \big ( x \, g_{\lambda_1}(0,-1;x)\big ){}'''$. Hence \freeW,
with \Ffree, reduces to
\eqn\sixexp{
{1\over x^4}  + C  + C'  {1\over (1-x)^4} 
= \sum_{\lambda_1} {\ts{ 1 \over (\lambda_1+1)\lambda_1(\lambda_1-1)}}\, 
a_{\lambda_1} \, \big ( x \, g_{\lambda_1}(0,-1;x)\big ){}''' \, ,
}
or
\eqn\sixee{
-{1\over x^2}  + C  x^2+ C'  {1\over x(1-x)}  + {a\over x} + b + c\, x
= \sum_{\lambda_1} {\ts{ 6 \over (\lambda_1+1)\lambda_1(\lambda_1-1)}}\, 
a_{\lambda_1} \,  g_{\lambda_1}(0,-1;x)  \, ,
}
with $a,b,c$ arbitrary. The solution is now
\eqn\twores{
a_{-2} = 1 \, , \qquad a_{p+2} = \big ( C(-1)^p + C' \big ) \, 
{(p+2)!\, (p+3)!\over 12(2p+1)!}\, , \quad p=0,1,\dots \, ,
}
and now  $F_{\! -2\,2} = F_{00} =1$.

\appendix{E}{Expressions in terms of Jack Polynomials}

Jack polynomials are a class of symmetric polynomials depending on
a parameter \Jack.
For two variables, which are sufficient for our purpose, we consider
$P^{(\vep)}_{\bl_1 \, \bl_2}(x,\bx)$, $\lambda_1 - 
\lambda_2 \in \Bbb N$, forming a basis for homogeneous symmetric functions in 
$x,\bx$ of degree $\lambda_1 + \lambda_2$. Moreover
\eqn\JP{ P^{(\vep)}_{\lambda_1 \lambda_2}(x,\bx) = (x\bx)^{\lambda_2} \, 
P^{(\vep)}_{\lambda_1-\lambda_2\, 0}(x,\bx) \, , } 
where $P^{(\vep)}_{\ell \, 0}(x,\bx)$ is a polynomial of degree $\ell$.
Here the normalisation is chosen so that 
\eqn\normJ{ P^{(\vep)}_{\bl_1 \, \bl_2}(x,x) = x^{\lambda_1+ \lambda_2} \, , 
\qquad P^{(\vep)}_{\bl_1 \, \bl_2}(x,\bx)  \limu{\bx\to 0,x\to 0} 
c^{(\vep)}_{\bl_1-\bl_2} x^{\bl_1}\bx^{\bl_2} \, , }
with $c^{(\vep)}_{\bl_1-\bl_2}$ given by \cell. For two variables 
Jack polynomials are expressible in terms of single variable 
Gegenbauer polynomials where with the conventions in \CCl
\eqn\JG{
P^{(\vep)}_{\lambda_1\lambda_2}(x,\bx) = 
(x\bx )^{{1\over 2}(\lambda_1+\lambda_2)} \, 
{\hat C}^{\, \vep}_{\lambda_1-\lambda_2 }(\si) \, ,
}
for $\si$ given by \defsi.

A formal solution for $F_{\bl_1 \bl_2}(x,\bx)$ was obtained previously \DT,
following \Koo, as a double series in terms of Jack polynomials 
$P^{(\vep)}_{\bl_1 \, \bl_2}(x,\bx)$ where 
\eqn\FJack{ F^{(\vep)}_{\bl_1 \bl_2}(x,\bx) = \sum_{{m,n \ge 
0\atop m-n + \lambda_1-\lambda_2 \ge 0}} r_{mn} \, P^{(\vep)}_{\bl_1+m \, 
\bl_2+n}(x,\bx) \, .} 
For $r_{00}=1$ \FJack\ satisfies \Fsim\ and indeed the first term in the 
expansion in \FJack\  is identical with the leading short distance 
contribution given by \Fxx\ and \fssol.

The coefficients $r_{mn}$ in \FJack\ are determined, starting from 
$r_{00}$, using
\eqn\DP{\eqalign{
\Delta^{(\vep)}(a,b)\, P^{(\vep)}_{\lambda_1\lambda_2} = {}& 
\big ( \lambda_1(\lambda_1-1) +
 \lambda_2(\lambda_2 -1 - 2\vep ) \big ) P^{(\vep)}_{\lambda_1\lambda_2} \cr
&{}- {\lambda_1 - \lambda_2 + 2\vep \over \lambda_1 - \lambda_2 + \vep} \,
(\lambda_1+a)(\lambda_1+b)\,P^{(\vep)}_{\lambda_1{+1}\,\lambda_2} \cr
&{} - {\lambda_1 - \lambda_2 \over \lambda_1 - \lambda_2 +  \vep} \,
(\lambda_2 +a - \vep)(\lambda_2 +b - \vep)\,
P^{(\vep)}_{\lambda_1\,\lambda_2{+1}} \, , }
}
which leads to a recursion relation for $r_{mn}$. Requiring $r_{mn}=0$
for $m-n+\lambda_1-\lambda_2=-1$ there is a general solution for any $\vep$ 
but it is very unwieldy, although the result was used in \DT\ to derive various 
recursion relations for $F^{(\vep)}_{\bl_1 \bl_2}(x,\bx)$ for arbitrary $\vep$.

If $\lambda_1 = \lambda_2 = \lambda$ the result simplifies to
\eqn\rsimp{
r_{mn} = {(\lambda+a)_m \, (\lambda+b)_m \, (\lambda+a- \vep)_n \, 
(\lambda+b- \vep)_n \over (2\lambda)_m \, (2\lambda - \vep)_n}\
{ (2 \vep)_{m-n} \, (m-n+  \vep)  \over n! \, (m-n)! \ 
( \vep)_{m+1}} \, ,
}
and if $\vep = - \half$ then
\eqn\rmmo{
r_{mn} = \cases{{(2\lambda+2a)_{2n} \, (2\lambda+2b)_{2n} \over
(2n)! \, (4 \lambda )_{2n} } \, , & $m=n  \, , $ \cr
{(2\lambda+2a)_{2n+1} \, (2\lambda+2b)_{2n+1} \over
(2n+1)! \, (4 \lambda )_{2n+1} }\, , & $m=n+1 \, ,$ \cr
0\, ,  & $m > n+1 \, . $}
}
It is easy to see that the summation reproduces \Fone, using \normJ.

When $\lambda_2 = \vep -b -N$, or similarly when $\lambda_2= \vep-a-N$, 
it is easy to see that in the solution \FJack\  the summation over $n$ 
truncates to $n=0,1,\dots, N$. 
Acting on Jack polynomials $ P^{(\vep)}_{\bl_1 \bl_2}(x,\bx)$ with the 
differential operator $\D^{(\vep)}$, defined by \defDD, gives
\eqn\DJP{
\D^{(\vep)} P^{(\vep)}_{\bl_1 \bl_2} = (\bl_1 + \vep) \bl_2 \, 
P^{(\vep)}_{\bl_1-1 \bl_2-1}  \quad \Rightarrow \quad \D^{(\vep)} 
P^{(\vep)}_{\ell\, 0} = 0 \, .
} 
Assuming \JG\ this is  equivalent to the standard differential equation for 
$C^{\vep}_\ell$. Applying \DJP\ to the solution \FJack\ for 
$(x\bx)^{-\lambda_2}F^{(\vep)}_{\lambda_1\lambda_2}(x,\bx)$, expressed in terms of 
$P^{(\vep)}_{\ell+m\, n}$, $n=0,1,\dots,N$, is an alternative justification
for  \DFh. For $N=0$ there is then just a single series so that 
\eqn\sinF{ F^{(\vep)}_{\bl_1 \, \vep-b}(x,\bx) = 
\sum_{m=0}^\infty \, {(\lambda_1+ a)_m \, 
(\lambda_1+b + \vep )_m\over m! \ (2\lambda_1)_m} \, 
P^{(\vep)}_{\bl_1+m \, \vep-b}(x,\bx) \, . }

For the case of primary interest here, $\vep=\half$, Gegenbauer polynomials 
become ordinary Legendre polynomials $P_\ell$ and Jack polynomials are
expressible just by
$P^{({1\over 2})}_{\ell\, 0}(x,\bx) = (x\bx)^{{1\over 2}\ell} P_\ell(\sigma) $
for $\sigma$ is defined in \defsi.
Using the representation provided by \Leg\ it is easy to see, using \JP, that
\sinF\ is identical with \thF\ and \Solfs\ since 
$c_\ell{}^{\!\smash{(1)}} = c_\ell$ in \cnorm. In general for $\vep=\half$
the Jack polynomial expansion \FJack\ can be recast as
\eqn\Jackh{
F^{({1\over 2})}_{\lambda_1\lambda_2}(x,\bx) = u^{\lambda_2} \, {1\over \pi}
\int_0^\pi \! \d s \; X^\ell
\sum_{{m,n \ge 0\atop m-n + \ell \ge 0}} r_{mn} \; 
\Big ( {u \over X} \Big )^n \, X^{m} \, , 
}
where we may note that, as a consequence of \fgred\ and \invx, 
for any integer $p$
\eqn\XXX{
( p +1) \, X^{p+1} + p \, u X^{p-1} - (p +\half) \, 
(x+\bx) X^p \sim 0 \, , \qquad X^p \sim {u^{p+{1\over 2}}\over 
X^{p+1}_{\vphantom d}} \, .
}
\XXX\ is equivalent to the standard recurrence relation for Legendre
polynomials.

\listrefs

\bye